\documentclass[aps,prb,twocolumn,superscriptaddress,showpacs]{revtex4}

\usepackage{graphicx}
\usepackage{dcolumn}
\usepackage{bm}

\begin{document}
\title{Competition of three-dimensional magnetic phases in Ca$_2$Ru$_{1-x}$Fe$_x$O$_4$: \protect\\  A structural perspective}

\author{Songxue Chi}
\affiliation{Neutron Scattering Division, Oak Ridge National Laboratory,
Oak Ridge, Tennessee 37831, USA}
\author{Feng Ye}
\affiliation{Neutron Scattering Division, Oak Ridge National Laboratory,
Oak Ridge, Tennessee 37831, USA}
\affiliation{Department of Physics and Astronomy,~University of Kentucky, Lexington, Kentucky 40506, USA}
\author{Gang Cao}
\affiliation{Department of Physics, University of Colorado at Boulder, Boulder, Colorado 80309, USA}
\author{Huibo Cao}
\affiliation{Neutron Scattering Division, Oak Ridge National Laboratory,
Oak Ridge, Tennessee 37831, USA}
\author{Jaime~A.~Fernandez-Baca}`
\affiliation{Neutron Scattering Division, Oak Ridge National Laboratory,
Oak Ridge, Tennessee 37831, USA}
\date{\today}

\begin{abstract}

The crystalline and magnetic structures of Ca$_2$Ru$_{1-x}$Fe$_x$O$_4$ (x=0.02, 0.05, 0.08 and 0.12) have been studied using neutron and X-ray diffraction. The Fe-doping reduces the Ru-O bond length in both apical and planar directions. The smaller Ru(Fe)O$_6$ octahedron leads to its reduced distortion. The $Pbca$ space group is maintained in all the Fe-dopings, so is the octahedral flattening. Warming has a similar effect on the lattice to that of the Fe-doping in releasing the distorted octahedra but precipitates an abrupt octahedral elongation near the N$\acute{e}$el temperature. Two competing antiferromagnetic orders, $A$- and $B$-centered phases have been observed. The Fe-doping-relaxed crystal structure prefers the latter to the former. As the doping increases, the $B$-centered phase continuously grows at the cost of the $A$-centered one and eventually replaces it at x=0.12. The absence of the two-dimensional antiferromagnetic critical fluctuations above the magnetic transition temperature and the three-dimensional magnetic correlation below the transition, together with the anomalous lattice response, point to an important role of orbital degree of freedom in driving the magnetic phase competition.

\end{abstract}

\maketitle

\begin{center}
${\bf I.~~INTRODUCTION}$
\end{center}

The single-layered perovskite Ca$_2$RuO$_4$ (CRO) encapsulates the most important themes underlining the latest trends in quantum materials research.  Bound to be metallic as its superconducting counterpart Sr$_2$RuO$_4$ with four 4$d$ electrons residing on six $t_{2g}$ bands, CRO is surprisingly insulating with a Curie-Weiss magnetic susceptibility \cite{Nakatsuji97} above its N$\acute{e}$el  temperature. It is widely accepted that its paramagnetic metal-insulator transition at $T_{MI}$ =357 K is a Mott transition \cite{Gorelov, Liebsch} caused by strong correlations.\cite{Hotta, Liebsch} Unlike the half-filled single-band 3$d$ electron systems such as cuprates and manganites whose Mott physics is a direct result of large Coulomb repulsion and small bandwidth, the nature of the insulating phase in CRO is more enigmatic and  has been a subject of continuing debate. The 2/3 filled $t_{2g}$ bands orbitals have a substantial orbital angular momentum, allowing Hund's rule coupling \cite{Sutter} and spin-orbit coupling (SOC) into the relevant energy range in forming the Mott gap. In the former case, Orbital-Selective Mott Transition (OSMT) \cite{Anisimov} was proposed which suggests that the Mott gap opens only on a subset of the $t_{2g}$ bands. \cite{Liebsch, Neupane, Gorelov, Pavarini04} Such a scenario later found its likely realization in iron pnictides. 
\cite{Haule, Werner, Medici, Chi} Alternatively as in the latter case, the MIT is suggested to be the product of both strong SOC and Mott physics, \cite{Liu11,Mizokawa} the combination of which triggers off a whole realm of quantum phenomena such as topological insulator, Weyl semimetals and quantum spin liquid. \cite{Witczak}

CRO becomes antiferromagnetically (AFM) ordered at $T_N$=112 K,\cite{Cao97, Nakatsuji97, Alexander, Braden1}, well below $T_{MI}$, suggesting its departure from the conventional
spin-only Mott magnetism. The consensus is that the unquenched SOC certainly plays a role,\cite{Nakatsuji2000,Mizokawa,Fatuzzo,Khaliullin, Kunkemoller15,Kunkemoller17,Jain,Zhang17}  but the importance of this role is where the controversies arise. One school of thoughts treat SOC as merely a perturbation for a local Hund's rule $S$ = 1 magnetic moment. \cite{Kunkemoller15,Kunkemoller17, Zhang17} Its opposing view suggests SOC is strong enough to bind local spin $S$ and orbital $L$ moments into a total angular momentum $j$, rendering a nonmagnetic $j$=0 ground state whose magnetic linear response function is of Van Vleck type.\cite{Akbari,Jain, Khaliullin} Inelastic neutron scattering \cite{Kunkemoller15,Kunkemoller17,Jain} revealed magnetic excitations that can be described by a conventional Heisenberg model and additional scattering features that cannot. Additionally, the soft amplitude mode of the spin-orbit condensate was observed by both INS \cite{Jain} and Raman scattering \cite{Souliou} measurements, which directly evidences excitonic magnetism. A recent RIXS study \cite {Das} reveals spin-orbital entangled excitations manifested within a band-Mott phase, reconciling the band-Mott and van Vleck-type Mott scenarios. 

The magnetic structure of CRO is $G$ type with propagation wave vector $\vec{q}$=(0,0,0). The ordered moment of the checkerboard-like AFM pattern is aligned along the orthorhombic $b$-axis \cite{Braden1}. Two types of nearest neighbor inter-plane arrangements have been found in the powdered sample:
Spins in one RuO$_2$ layer simply
shift from the next layer by (0, b/2, c/2) or (a/2, 0, c/2), which were called $A$-centered and $B$-centered respectively.\cite{Braden1}
These two magnetic phases can coexist and their relative proportion can be tuned by 
oxygen content, \cite{Braden1}, pressure, \cite{Steffens05} and chemical doping. \cite{Friedt, Pincini18} However, the nature of such magnetic competition has yet to be determined.
As the structural carrier of such rich interplays among spin, charge and orbital degrees of freedom, RuO$_6$ octahedron becomes an effective control knob \cite{Jung, Liu11, Snow, Hotta, Lee02, Qi_Cr, Qi_Fe} through its crucial role in SOC\cite{Fatuzzo}, crystal field splitting\cite{Braden1}, Jahn Teller coupling\cite{Liu19} and spin-phonon coupling\cite{Liu19,Lee19}. The distortions of the octahedra lie at the heart of electronic and magnetic phenomenology in CRO and its derived compounds. A more adequate physical interpretation of the aforementioned issues demands a simultaneous tracing of the magnetic phases and the octahedral structure.

In this paper, we report an investigation of CRO using both neutron and X-ray diffraction that details the structural and magnetic changes caused by Fe-substitution for Ru ions in CRO. The results of this investigation show that the increased Fe doping releases the distorted RuO$_6$ octahedra. With the unchanged $Pbca$ symmetry, the unbuckled octahedra are accompanied by a systematic transition between two AFM phases, which compete and result in short-range magnetic orders although their three-dimensional character remains. Our findings highlight the vital role played by the spin orbital correlation in determining the magnetic ground state and demonstrate how the RuO$_6$ octahedra can tune it.
The scheme of the paper is the following. In Sec. III we present the details of the crystal structures at 240 K for four Fe-concentrations, and also the temperature dependence of the octahedral deformation in the x=0.08 system. Then we show the results of the magnetic diffraction measurements including the spin arrangements
of the magnetic phases in the four compounds, their evolution with temperature. Lastly we present the in-plane and out-of-plane magnetic correlation lengths 
as well as the critical behavior close to $T_N$ in x=0.08. In Sec. IV, we discuss the influence of the structural evolution on the magnetic phase competition.

\begin{figure}
\label{fig1}
\includegraphics [width=1.0\columnwidth]{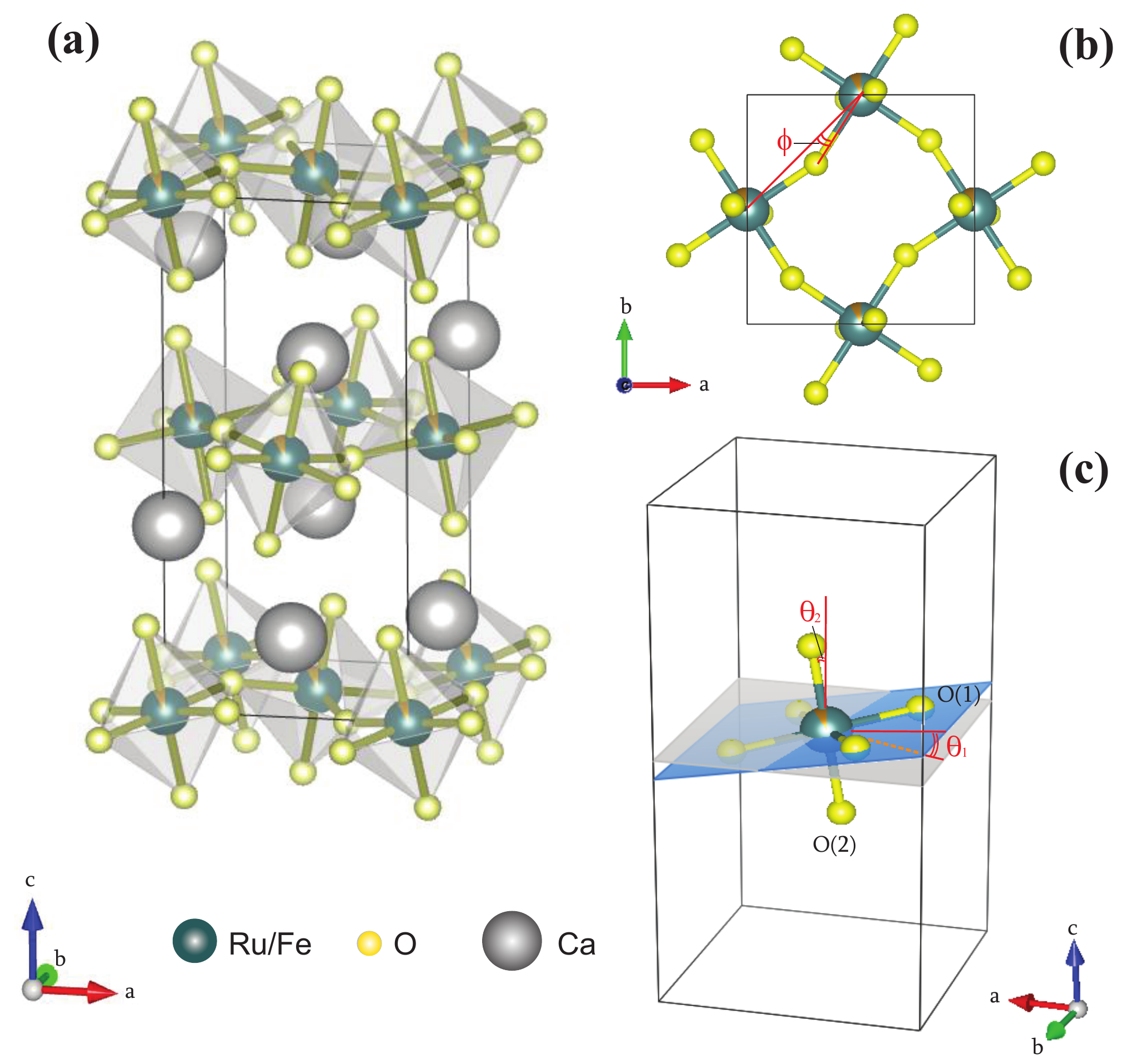}
\caption{(color online) (a) Crystal structure of Ca$_2$RuO$_4$. The gray spheres represent calcium atoms, the green spheres ruthenium atoms and the yellow ones oxygen atoms. (b) The top view of the RuO$_2$ plane shows the rotated oxygen octahedra. The rotation angle $\phi$ is defined as the projection of the angle between the Ru-O(2) bond and the Ru-Ru bond onto the $ab$-plane. (c) The tilt angle $\theta$ is defined as the angle between the O(2) plane and the $ab$-plane.
}
\end{figure}

\begin{center}
${\bf II. ~~EXPERIMENTAL}$
\end{center}

Single crystals of Ca$_2$Ru$_{1-x}$Fe$_x$O$_4$ with x=0.02, 0.05, 0.08 and 0.12 were grown by the floating-zone method, the details of which are described elsewhere. \cite{Qi_Cr} The crystals were mounted in closed cycle refrigerators for neutron diffraction measurements on HB2C wide angle neutron diffractometer (WAND) and HB3A four-circle diffractometer, as well as triple-axis spectrometers HB3 and HB1A at the High Flux Isotope Reactor at Oak Ridge National Laboratory (ORNL). The wavelengths were 1.482 $\text{\AA}$ for HB2C, 1.003 $\text{\AA}$ and 1.542 $\text{\AA}$ for HB3A, and 2.359 $\text{\AA}$ for HB3 and HB1A. The collimations of 48'-40'-sample-40'-120' were used for HB3 measurements, and 40'-40'-sample-40'-80' for HB1A. X-ray diffraction data were collected using a
Rigaku XtaLAB PRO diffractometer with a Dectris Pilatus 200K detector. A molybdenum anode was used to generate X-ray with wavelength $\lambda$ = 0.7107 $\text{\AA}$. The samples were cooled by cold nitrogen flow provided by an Oxford N-HeliX Cryosystem. Empirical absorption correction was applied in the process of data collection, which was integrated and scaled using the CrysAlisPro. The X-ray structure data were solved and refined using SIR-2011 in WinGX and SHELXL-2013
software packages. \cite{SHELXT, Farrugia, Burla} The representation analysis and structure refinement with neutron data were conducted using FullProf Suite. \cite{FullProf}Positions in reciprocal space are given in reduced lattice notation where $\vec{Q}[r.l.u.]=\vec{Q}[\text{\AA}^{-1}].(\frac{a}{2\pi}\hat{h}+\frac{b}{2\pi}\hat{k}+\frac{c}{2\pi}\hat{l})$.

\begin{center}
{\bf III. ~~RESULTS} 
\end{center}
 
\begin{center}
{\bf A. Doping dependence of the crystal structure}
\end{center}

The structure of Ca$_2$RuO$_4$ and its derivatives can be understood through its deviation from the ideal K$_2$NiF$_4$ structure which has 
the space group $I4/mmm$. In the high-temperature tetragonal phase, the corner-sharing RuO$_6$ octahedra have a staggered correlated rotation about the long axis $c$. In the low-temperature insulating phase the octahedra are further distorted, as depicted in Fig. 1(a). Besides the rotation about $c$, such distortion also involves a tilt of the octahedra about an axis in the $ab$-plane. This tilt axis is the line of intersection of the $ab$-plane and the basal oxygen plane of the RuO$_6$ octahedra. The rotation angle shown in Fig. 1(b) is the angle between the octahedron tilt axis and the edge of an octahedron basal 
plane is the projection of the actual rotation to the $ab$-plane. The four oxygen atoms comprising this basal plane are derived from one oxygen site permitted by the $Pbca$ symmetry, although there are two Ru-O(1) distances. We label the rotation angle $\phi$ and the tilt angle $\theta$ to be consistent with the structural report on the parent compound. \cite{Braden1}. Our structural characterizations of its Fe-doped derivatives were carried out by both neutron and X-ray diffraction, which agree on the group symmetry and the trend of structural transformation. Our discussion on the structural details is based on the X-ray data while the temperature variations of the lattice constants on the neutron data.  Using the space group determination module, GRAL, on CrysAlisPro we confirm that the crystals of all four Fe-concentrations retain the same $Pbca$ space group. The experimental and refinement details of the single-crystal X-ray diffraction are given in Table 1.

\begin{table*}[ht!]
\caption{Structural refinement details of Ca$_2$Ru$_{1-x}$Fe$_x$O$_4$ with the data obtained using a Rigaku XtaLAB PRO diffractometer 
}
\label{tab1}
\begin{ruledtabular}
\begin{tabular}{llccccc}
 x & x=0.02 & x=0.08  & x=0.08 & x=0.08 & x=0.12  \\
     T  & 240 K & 100 K & 150 K & 240 K  & 240  \\
		 \hline\hline
				
  $a$ & 5.4082(1) & 5.3920(1)& 5.3994(8) &5.3989(1)&5.3989(1) \\
	
	$b$ & 5.5614(1) & 5.6111(1) & 5.6073(3) &5.5370(3)&5.5370(1)  \\

  $c$ & 11.8437(3) & 11.7404(3) & 11.7567(8) & 11.8283(3) & 11.8283(3)  \\
	
	Vol ($\text{AA}$$^3$) & 355.21(0) & 355.20(1) & 355.95(3) & 353.59(0) & 353.59(1) \\

  $Ru/Fe$    &  0.500(4)  &  0.500(2) &  0.500(3)   &  0.500(4)  & 0.500(4)  \\

  $Ca(x)$  & 0.4938((6) & 0.4958(4) & 0.4957(4) & 0.4933(6) & 0.4924(6) \\
	
  $Ca(y)$  & -0.0519((2) & -0.0576(3) & -0.0572(3)& -0.0509(2) & -0.0489(4) \\

  $Ca(z)$  & 0.3518(3) & 0.3524(0) & 0.3524(1) & 0.3517(2) & 0.3515(2) \\
	
  $O1(x)$ & 0.1968(9) & 0.1952(6) & 0.1955(7)  & 0.1970(9) & 0.1977(9)  \\

  $O1(y)$ & 0.1994(8) & 0.1989(5) & 0.1993(3) & 0.1995(7) & 0.1997(1) \\

  $O1(z)$ & 0.0260(9)& 0.0276(4)  & 0.0274(6) & 0.0258(5) & 0.0250(4) \\

  $O2(x)$ & 0.5646(5) & 0.5689(4) & 0.5685(4)  & 0.5642(5) & 0.5616(4)  \\

  $O2(y)$ & 0.0197(6) & 0.0216(1) & 0.0211(1) & 0.0194(2) & 0.0188(4) \\

  $O2(z)$ & 0.1648(9) & 0.1647(6)  & 0.1645(3) & 0.1649(7) & 0.1648(1) \\
	
	No. of measured reflections & 3364 & 3298  & 3965 & 7756 & 7830 \\
	
	No. of unique reflections & 510 & 447  & 588 & 674 & 622 \\
	
	No. of observed [$I$$>$2$\sigma$($F^2$)]    &    &  405  & 628   & 562   &   \\
	
	reflections & 465 & 447 & 610 & 674 & 622 \\
	
  No. of parameters & 36 & 36 & 36 & 36 & 36 \\
		
  Goodness-of-fit on F$^2$& 1.182 & 1.130  & 1.201 & 1.172 & 1.074 \\
	
	wR2 & 0.0381 & 0.0297  & 0.0372 & 0.0282 & 0.0862 \\
	
	R1 [I$>$2$\sigma$(I)] & 0.0137  & 0.0111  & 0.0135 & 0.0106 & 0.0325 \\

	R1 for all & 0.0152  & 0.0130 & 0.0139 & 0.0.0120 & 0.0342 \\

\end{tabular}
\end{ruledtabular}
\end{table*}

The structural parameters that describe the shape of the RuO$_6$ octahedron are plotted in Figure 2 to illustrate 
the changes brought about by the Fe-substitution.
The comparisons are made at 240 K for three Fe-concentrations: x=0.02, 0.08 and 0.12. The four parameters, the Ru-O bond lengths, the Ru-O-Ru bond angle, the in-plane rotation angle and the tilt angles are shown. Both Ru-O(1) and Ru-O(2) bond lengths decrease as doping increases because of reduced ion size on the $4a$-site. 
The shrinkage of the octahedron by Fe-substitution preserves its apical flattening though, in which the Ru-O(1) bond remains slightly longer than the Ru-O(2) bond. The ratio of apical bond length Ru-O(2) to the basal Ru-O(1), $\delta$, is increased by the added Fe but remains below 1 for all dopings as shown in the inset of Fig. 2(a). The x=0.08 system has the greatest value of this ratio which is 0.9926(7). We discuss the significance of this ratio in the Discussion section. The reduction of the octahedron volume brought by Fe-doping tends to release its buckled position and thus reduces the orthorhombic strain in general.  Figure 2(c) and 2(d) show that both the planar rotation $\phi$ and the tilt $\theta$ decrease as Fe concentration increases. The symmetry allows 
independent tilts of the basal O(1)-plane and the apical O(2) axis. The basal projection of the apical O(2) tilt deviates from the $a$-axis by an angle that is very close to $\phi$ which is the basal rotation. The Fe substitution reduces both tilts. The apical tilt remains greater than that of the basal plane by the same amount for all the Fe concentrations as the case in the pristine CRO. The aforementioned details of the bond distance and tilt angles indicate that Fe-substitution reduces the positional distortion of the RuO$_6$ octahedron but maintains its compressed shape. As a result of the reduced octahedral rotation and tilt, the Ru-O-Ru bond angle decreases with increasing Fe-doping as shown in Fig. 2(b).

\begin{figure}
\label{fig2}
\includegraphics [width=1.0\columnwidth]{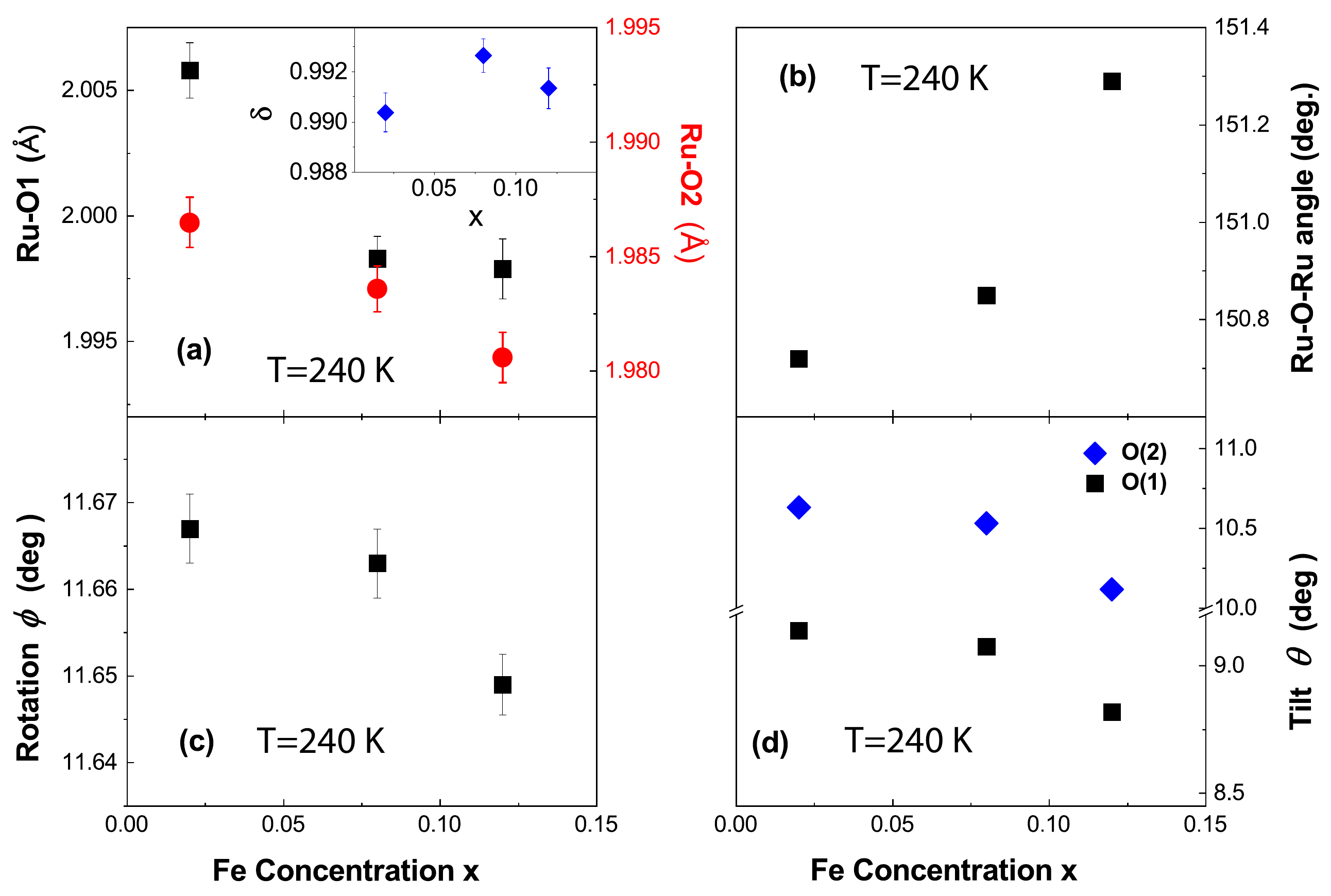}
\caption{(color online)  Fe-concentration dependence of the (a) Ru-O(1) and Ru-O(2) bond length, (b) Ru-O-Ru bond angle, (c) rotation angle of the Ru-O$_6$ octahedra and (d) tile angle, $\theta$, of the basal O(1) plane and that of the apical O(2) at 240 K. The inset in (a) shows the ratio $\delta$ of apical bond length Ru-O(2) to planar bond length Ru-O(1) that represent the octahedral flattening.
}
\end{figure}

\begin{figure}
\label{fig3}
\includegraphics [width=1.0\columnwidth]{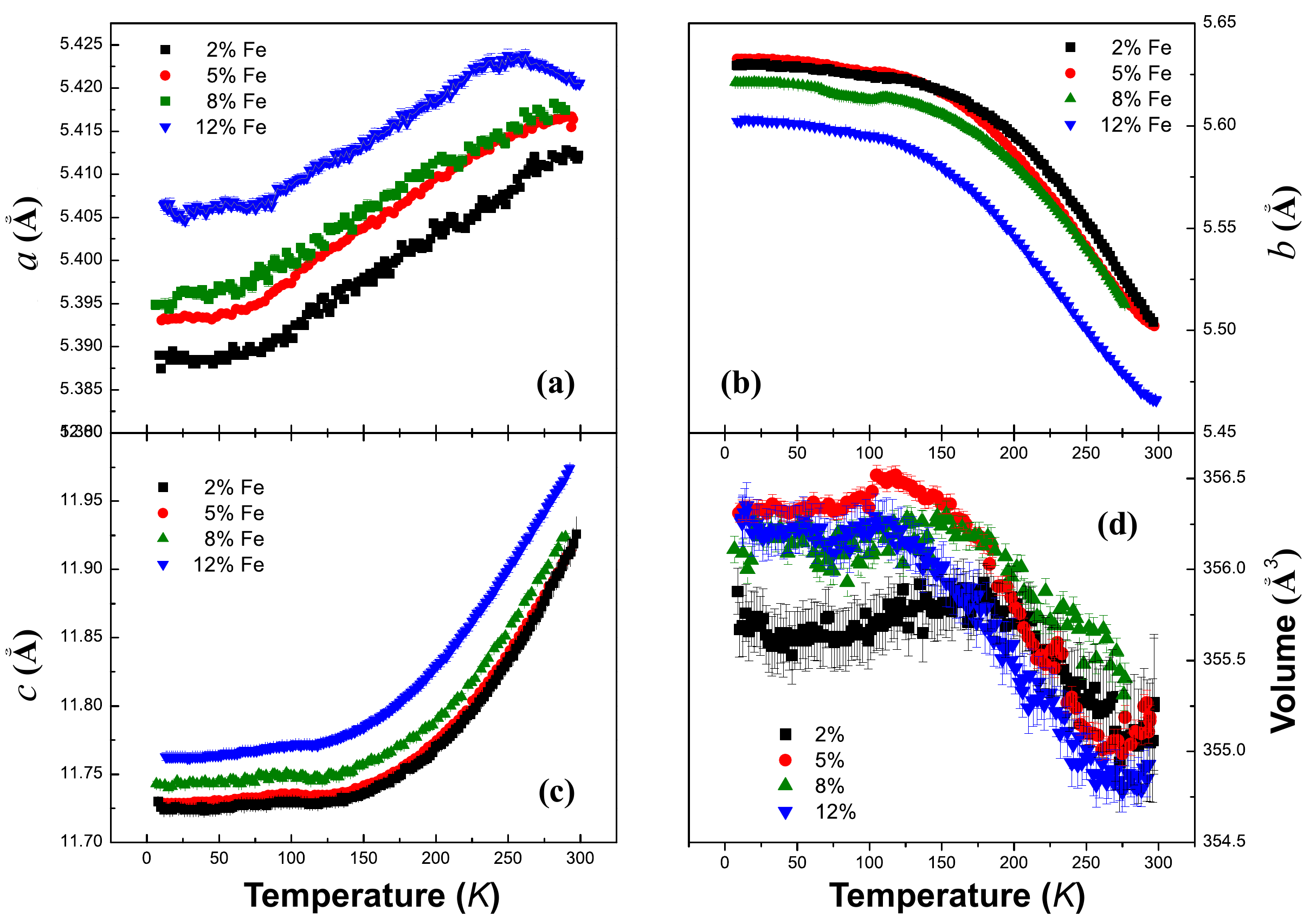}
\caption{(color online) The temperature dependence of lattice constants and volumes in Ca$_2$Ru$_{1-x}$Fe$_x$O$_4$.
}
\end{figure}

\begin{center}
{\bf B. Temperature effect on the crystal structure }
\end{center}

We extract the lattice constants from the $\theta$-2$\theta$ scans on well aligned single crystals with Triple-axis neutron spectrometers. The temperature dependence of the lattice constants of the four Fe-concentrations are summarized in Figure 3.  At any given temperature in the measured range from 4 K to 300 K lattice constant $a$ and $c$, as shown in Fig. 3(a) and Fig. 3(c), increases as Fe concentration increases. Lattice parameter $b$ as in Fig. 3(b) shows the opposite trend: the Fe-substitution shortens $b$. The trend that Fe-doping reduces orthohombicity stays true throughout the entire temperature range we measured (4 K to 300 K). For each doping, cooling enhances orthohombicity as in the parent CRO compound. 

The most rapid variation of all lattice constants occur at high temperatures and a saturation sets in at low temperatures. Although the variation of the lattice constants is generally continuous in the measured temperature range, $b$ and $c$ clearly show anomalous change around $T_N$, where abrupt shortening of $b$ and increase of $c$ are visible. These lattice anomalies are most pronounced in x=0.05 and 0.08 as shown by the red circle and green square in Fig. 3(b) and 3(c).
The volume of the unit cell, as a result of the combined effect, exhibits negative volume thermal expansion (NVTE) as shown in Fig. 3(d), which is consistent with the previous report \cite{Qi_Fe} and similar to Mn-\cite{Qi_Fe} Cr-doped CRO. \cite{Qi_Cr}. For x=0.02 NVTE gradually disappears below 200 K and become independent of temperature, while the three higher dopings see a clear transition close to $T_N$ that separates NVTE and and an Invar-like thermal response. 
Our temperatures do not reach $T_{MI}$ ($>$350 K). We know from the previous work that the increased doping increases $T_{MI}$ first then MIT gets smeared at higher dopings. \cite{Yuan} The drop of the lattice parameter $a$  of
x=0.12 around 250 K on warming may hint the reduced MIT transition temperature.

We chose x=0.08 for a detailed X-ray diffraction study at a few representative temperatures to learn the evolution of the octahedral distortion, as summarized in Fig. 4, its correlation with the lattice constants and with the magnetic phases.  In the pristine CRO, the flattened RuO$_6$ octahedron undergoes further compression along the apical direction on cooling. The decrease of Ru-O(2) and increase of Ru-O(1) are continuous until the saturation is reached at low temperatures. \cite{Braden1} The Fe-doped x=0.08 system shows similar trend above T$_N$, as shown in Fig. 4(a). However, such tendency is reversed as the system cools below T$_N$, where Ru-O(1) exhibits abrupt decrease and Ru-O(2) increase at 100 K. The value of $\delta$ remains less than 1 though, which means the octahedron becomes less flattened but elongation has not occurred. The general effect of cooling is to enhance the octahedral distortion.  Ru-O-Ru bond angle is reduced by cooling. The octahedral rotation $\phi$ and the tilt $\theta$, including apical O(2) and basal O(1), increase as displayed in Fig. 4(c) and 4(d).  $\phi$ in the parent CRO compound is almost temperature independent in this range.\cite{Braden1} The Fe-substitution frees the space to allow the octahedral rotation. The increase of $\theta$ is responsible for the increased orthorhombicity, namely the shortened $a$ and enhanced $b$, at low temperatures. Given the smooth changes of rotation and tilt over the temperature, it is clear that abrupt bond length changes are responsible for the anomalies in the temperature variation of lattice $b$ and $c$ across the magnetic transition in Fig. 3(b) and 3(c).

\begin{figure}
\label{fig4}
\includegraphics [width=1.0\columnwidth]{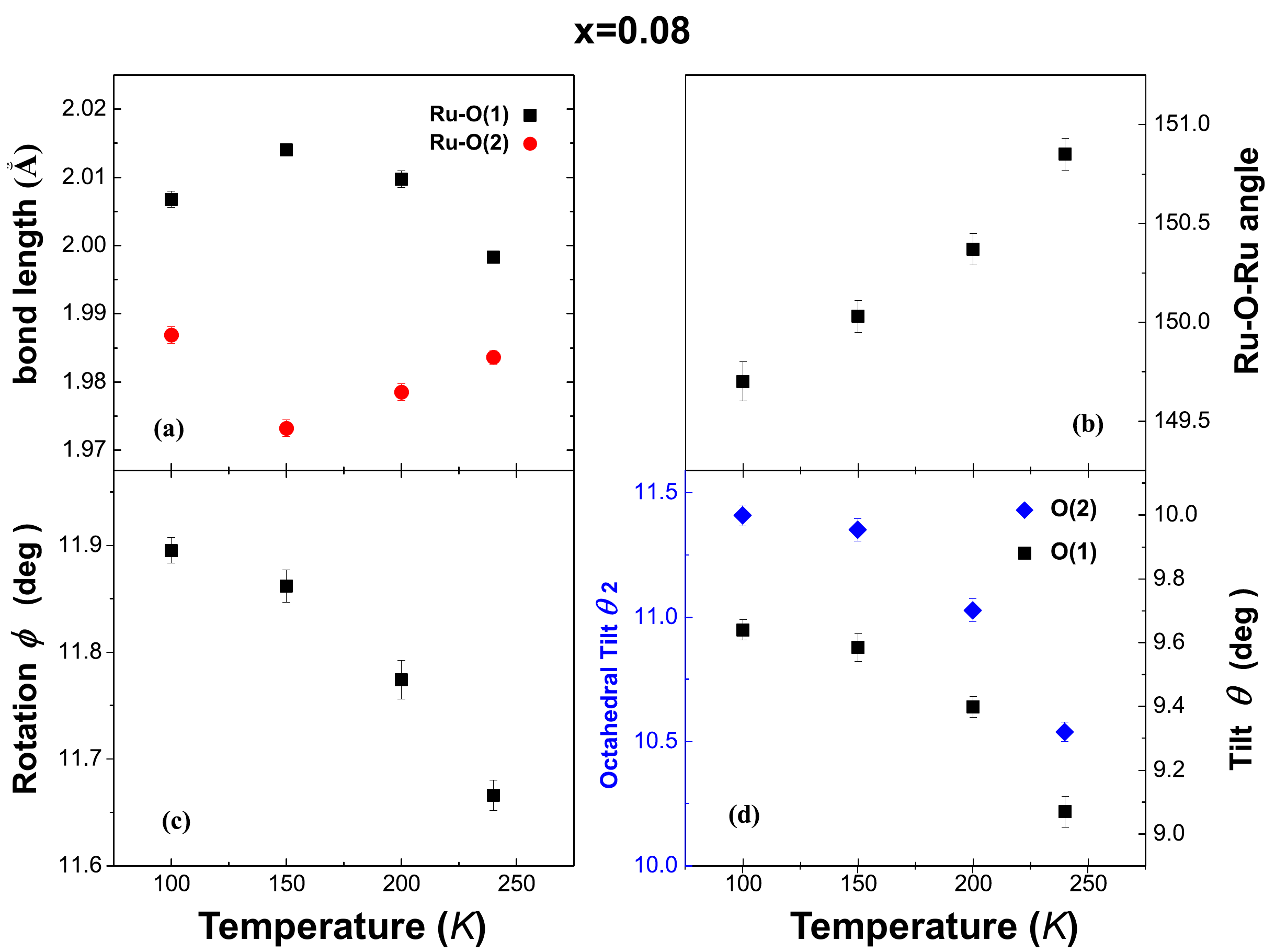}
\caption{(color online)  The temperature dependence of the octahedral distortion in x=0.08. (a) The Ru-O(1) (black square) and Ru-O(2) (red circle) bond lengths, (b) the Ru-O-Ru bond angle, (c) the rotation angle $\phi$ of the Ru(Fe)O$_6$ octahedron, (d) the tilt $\theta$ of the inplane O(1) (black square) and the apical O(2) (green diamond) as a function of temperature.}.
\end{figure}

\begin{center}
{\bf C. The spin structures of the two magnetic phases}
\end{center}

The magnetic structures of the two phases in the parent CRO \cite{Braden1} have identical in-plane spin arrangement. Their difference lies in the inter-layer arrangements which is highlighted by the shaded plane in Figure 5.

\begin{figure}
\label{fig5}
\includegraphics [width=1.0\columnwidth]{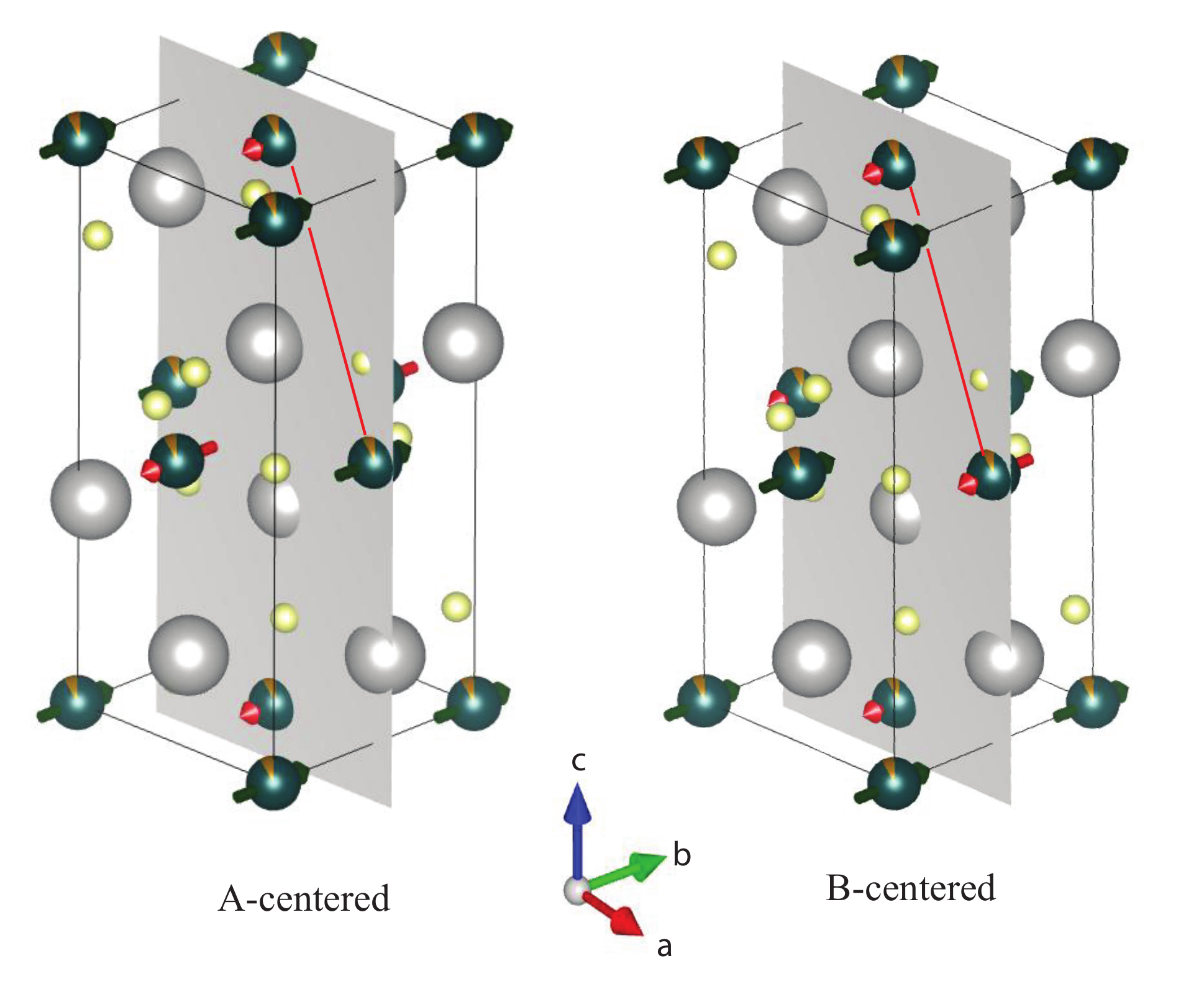}
\caption{(color online) Two stacking schemes for the Ru$^{4+}$ ordered moments which lie along the $b$-axis of orthorhombic lattice. The ferromagnetically correlated moments lie in the planes that are parallel to the (1,0,0) plane in the case of A-center and parallel to the (0,1,0) plane in B-centered case.
}
\end{figure}

The intensity of a magnetic Bragg peak is proportional to  
$
{|{F_M}(\bf{q})|^2}/{\sin(2\theta)}
$, where $\theta$ is the
scattering angle and ${F_M}$ the magnetic structure factor which is given by \cite{Blume}
\begin{equation}
F_M({\bf{q}}) = \sum_{j}f({\bf{q}})_j \left\langle \mu_z \right\rangle e^{i {\bf{q}} \cdot {\bf{r}}
} e^{-W_j},
\end{equation} where $f(\bf{q})_j$, $\left\langle \mu_z \right\rangle$ and $e^{-W_j}$ are the magnetic
form factor, the thermal average of the aligned magnetic moment of the $j$-th ion and Debye-Waller factor 
respectively. For a system with magnetic moments of the same type, this expression simplifies to
\begin{equation}
F_M({\bf{q}}) = f({\bf{q}})_j \left\langle \mu_z \right\rangle  e^{-W_j}\sum_{j}e^{i {\bf{q}} \cdot {\bf{r}}},
\end{equation}

Because of the orthorhombic structure, 
these two arrangements are not geometrically equivalent. At 100 K, the interlayer nearest neighbor (NN) distance is 6.4823(14) $\text{\AA}$ and next nearest neighbor (NNN) distance is 6.52416(14) $\text{\AA}$, so the magnetic exchange is
not frustrated. The NN interaction, as indicated by the red lines in Fig. 5, is AFM for the $A$-centered phase and FM for the $B$-phase.  Executing the summation of Eq.(2) on the Wyckoff Position 4a yields the integrated intensity of a magnetic reflection

\begin{equation}
I \propto \left|f({\bf{q}})\right|^2 \left|1-e^{i \pi (H+K)}\pm e^{i \pi (K+L)}\mp e^{i \pi (H+L)}\right|^2
\end{equation}
where the $\pm$ signs are for $A$- and $B$-centered inter-layer arrangements respectively.

For the $A$-centered phase, the parity of $H$ should be different from that of the $K$ and $L$ in order to have non-zero value of the magnetic structure factor. This condition for the $B$-centered case requires the parity of K to be different from that of H and L.   Following the conditions dictated by equation (3), the expected magnetic reflections in these two planes are $(2m+1,0,2n)/(0,2m+1,2n+1)$ for $A$-centered phase, and $(2m+1,0,2n+1)/(0,2m+1,2n)$ for the $B$-centered, where m and n are integers. 

To trace the effect of Fe-substitution on the magnetic orders, we started by surveying the scattering planes $(H,K,0)$, $(H,0,L)$ and $(0,K,L)$ of various Fe-dopings with the WAND diffractometer taking advantage of its wide angle detector. The room temperature contour plots in the $(H,K,0)$ plane (data not shown) exhibit not extra peaks for the the single crystals  of all the 4 Fe-dopings, indicating they have single domain. In x=0.02, reflections disallowed by the $Pbca$ symmetry were found at positions such as (1,0,0) and (0,1,1) at low temperatures, which meet the conditions for an $A$-centered magnetic phase. The temperature dependence of the (1,0,0) position shows the disappearance of its intensity above 113 K, as shown in Fig. 8(a), confirming its magnetic origin. Although its N$\acute{e}$el temperature agrees with that of the parent compound, x=0.02 compound exhibit no trace of a co-existing $B$-centered phase as in its polycrystalline parent. \cite{Braden1} It is unlikely that the slight Fe-substitution suppresses 
the co-existing phase because there has been report of achieving a single magnetic phase in high quality Ca$_2$RuO$_4$ single crystal.\cite{Steffens05}

Magnetic peaks attributed to the $B$-phase, such as (1,0,1) and (0,1,2), appear in the x=0.05 compound, along with the $A$-phase peaks.  As Fe-content increases the $B$-phase grows at the expense of the $A$-phase which is evidenced by the change of their relative intensities in x=0.08.  Figure 6 shows examples of the coexistence of these two phases in x=0.08 at 4 K and 140 K in the $(H,0,L)$ and $(0,K,L)$ planes . In the reciprocal-space map of the $(H,0,L)$ plane, reflections from the two magnetic phases are both visible at 4 K [Fig.6(a)]. They include peaks with even-number $L$ ($A$-phase) and odd-number $L$ ($B$-phase) in the [1,0,$L$] direction which disappear at 140 K [Fig.6(b)]. Similarly in the [0,1,L] direction of the $(0,K,L)$ plane, as in Fig.6(c), peaks with odd- and even-number $L$ are both visible at 4 K and both vanish at 140 K as in Fig.6(d). Figure 7 displays cuts along the [1,0,$L$] direction from the $(H,0,L)$ map at various temperatures. The elevated temperature has different effects on the two sets of magnetic peaks. Heating from 10 K to 50 K seems to have very little effect on all the peaks. However at 95 K, all peaks with even $L$ disappeared while all those with odd $L$ gain substantial intensities before disappearing at 113 K.

More detailed temperature dependence measurements, as shown in Fig. 8, reveal the competing nature of the two magnetic phases and that the Fe-substitution
prefers one to the other. The intensities of the magnetic peaks (1,0,0) and (1,0,1) in the compounds of different Fe-dopings have been normalized using their nuclear peak intensities. In the x=0.05 compound [Fig. 8(b)], the (1,0,0) peak is a bit weaker than that in x=0.02 [Fig. 8(a)] at 4 K. The $B$-phase peak (1,0,1) appears and its remains almost unchanged as temperature rises until the intensity of (1,0,0) starts to decrease. (1,0,1) gradually intensifies, as (1,0,0) weakens, and reaches its maximum value at about 105 K before disappearing concomitantly with (1,0,1) at 113 K. At x=0.08, as shown by the normalized intensity in Fig. 8(c), (1,0,1) continues to grow in intensity while (1,0,0) weakens. Again the $A$-centered phase prevails in the competition at elevated temperatures, only this time completely suppressed the (1,0,1) peak below 80 K. The transition temperature of the A-phase remains the same at 113 K.
As the Fe concentration reaches 0.12, the (1,0,0) peak is vanquished and (1,0,1) exhibits an undisturbed order parameter with a unchanged transition temperature. 
 
\begin{figure}
\label{fig6}
\includegraphics [width=1.0\columnwidth]{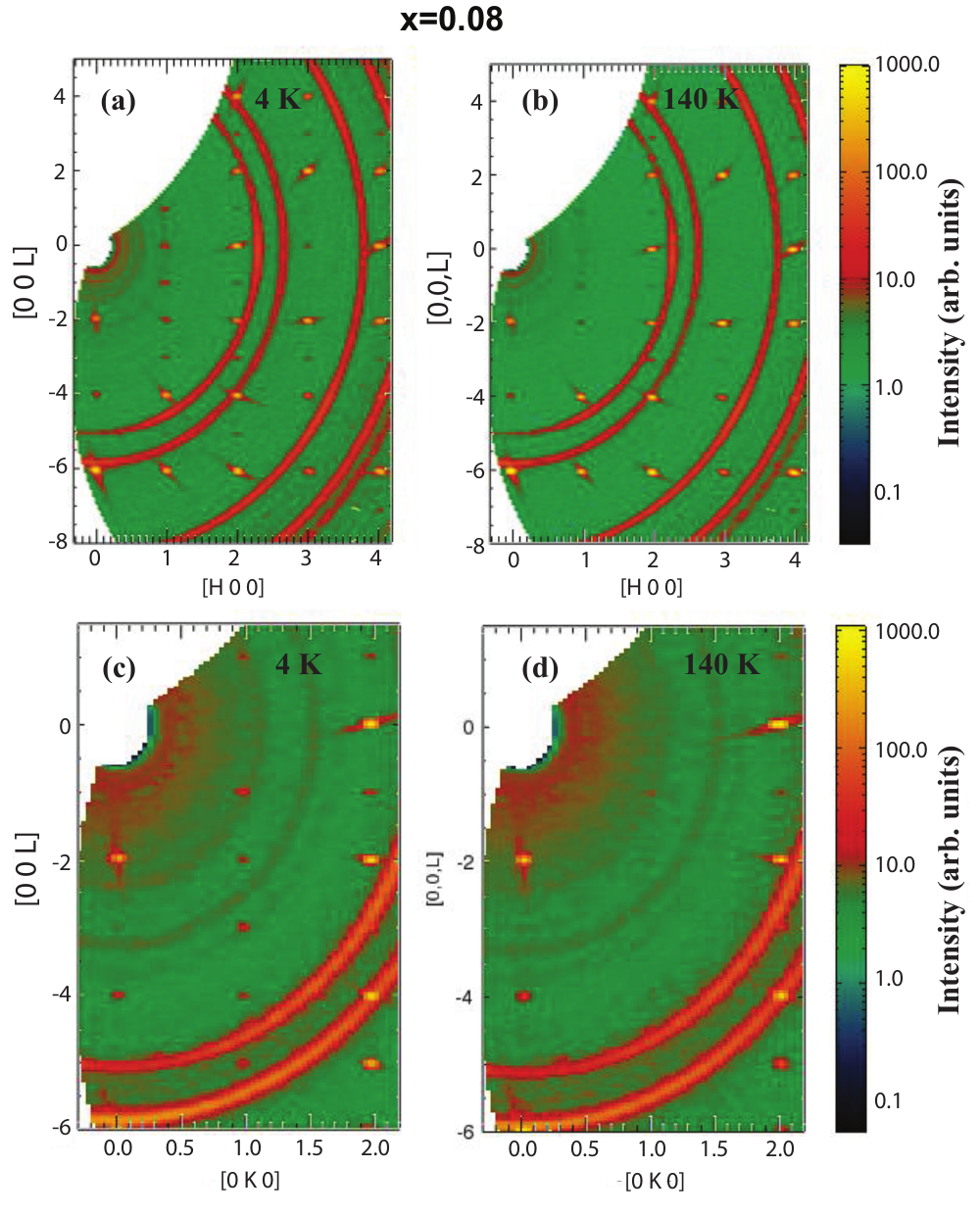}
\caption{(color online) The contour plot of the diffraction data collected on WAND in the (H,0,L) scattering plane for x=0.08 at (a) 4 K and (b) 140 K. (c) and (d) show the contour maps of the (0,K,L) scattering plane for the same sample at 4 K and 140 K respectively. 
}
\end{figure}

\begin{figure}
\label{fig7}
\includegraphics [width=1.0\columnwidth]{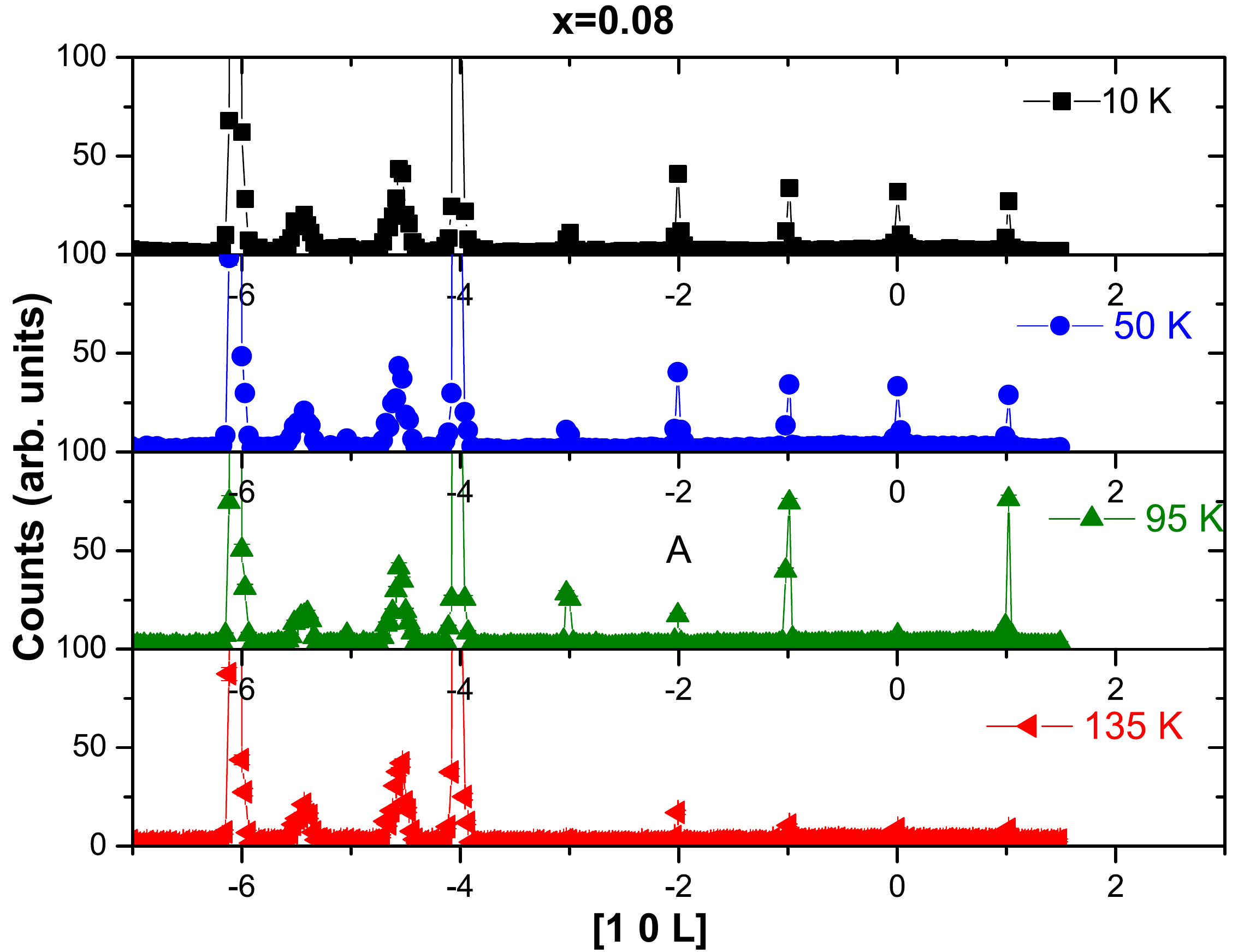}
\caption{(color online) Line cuts of the WAND data for x=0.08 along the [1,0,L] direction at (a) 10 K, (b) 50 K, (c) 95 K and (d) 135 K that show different temperature dependencies of peaks with odd and even $L$ values.
}
\end{figure}

To precisely characterize the spin configurations, we collected magnetic as well as nuclear reflections for single crystal of each composition using the 4-circle diffractometer HB3A. Their integrated intensities were used for structural refinement using FullProf program suite. The crystal structure refinement described in the previous section provides the scale factor, extinction parameter, atomic parameters including positions and thermal displacement parameters. The magnetic intensities, obtained from subtracting the high-temperature intensity, were used to refine the orientation and the size of the ordered moment. Representation analysis provides four different irreducible representations (irreps) $\Gamma_1$, $\Gamma_3$, $\Gamma_5$ and $\Gamma_7$, each of which consists of three basis vectors. We sorted through all basis vectors and their combinations for each irreps for each set of magnetic peaks. The Rietveld refinement reached convergence with only $\Gamma_1$ and $\Gamma_3$ which correspond to the $A$- and the $B$-centered spin structures respectively. The magnetic space group for $\Gamma_1$ is $Pbca$ (BNS: 61.433; OG : 61.1.497) and that for $\Gamma_3$ is $Pb'c'a$ (BNS: 61.436; OG: 61.4.500). Here BNS and OG refer to the Belov-Neronova-Smirnova notation \cite{BNS} and the Opechowski-Guccione notation,\cite{OG} respectively. The best R-factors were obtained when ordered moment lie along the $b$-axis without any measurable staggered moment along $a$ or $c$. The summary of the refinements for all four compounds is tabulated in Table II. Ru has an intermediate spin configuration $t_{2g}^4e_g^0$. A CEF splitting between $t_{2g}$ and $e_g$ shells stabilizes this spin state whose full moment is 2$\mu_B$. The total ordered moment in the pristine CRO compound was determined to be 1.3 $\mu_B$. \cite{Braden1} The reduction of the ordered moment in the parent compound is believed to be caused by the strong co-valency between the Ru 4$d$ and O 2$p$ orbitals. Its total angular momentum may also be reduced by spin and orbital fluctuations. \cite{Mizokawa} The Fe-substitution generally increases the ordered moment, which is found to be 1.62(7) $\mu_B$ in the x=0.02 crystal. As the Fe content increases and the $B$-phase grows, the ordered moment in the $A$-phase decreases. However, the summed moment of the two phases decreases in x=0.05 and 0.08. As the $B$-phase completely takes over in x=0.12, the moment shows some recovery at 1.38(7) $\mu_B$.

The fact that the refined moment points to the $b$-axis in both $\Gamma_1$ and $\Gamma_3$ structures rules out the scenario where a gradual deviation of the ordered moment from the $b$-axis induced by Fe-doping. In such a picture, the ordered moment develops a component in the $a$-direction so that the magnetic structure factor always has non-zero components in Eq. (2) regardless of the spin orientations. Another important result in the x=0.08 system is that the two magnetic sub-lattices exhibit different N$\acute{e}$el temperatures, which can rule out the possibility of a quantum superposition of the two AFM configuration by removing the local orbital quenching in the angular momentum. \cite{Cuoco2} Our results on the magnetic structure and on the evolution of the magnetic correlation indicate that the two sets of magnetic reflections belong to separate magnetic phases that compete for the same lattice. 

\begin{figure}
\label{fig8}
\includegraphics [width=1.0\columnwidth]{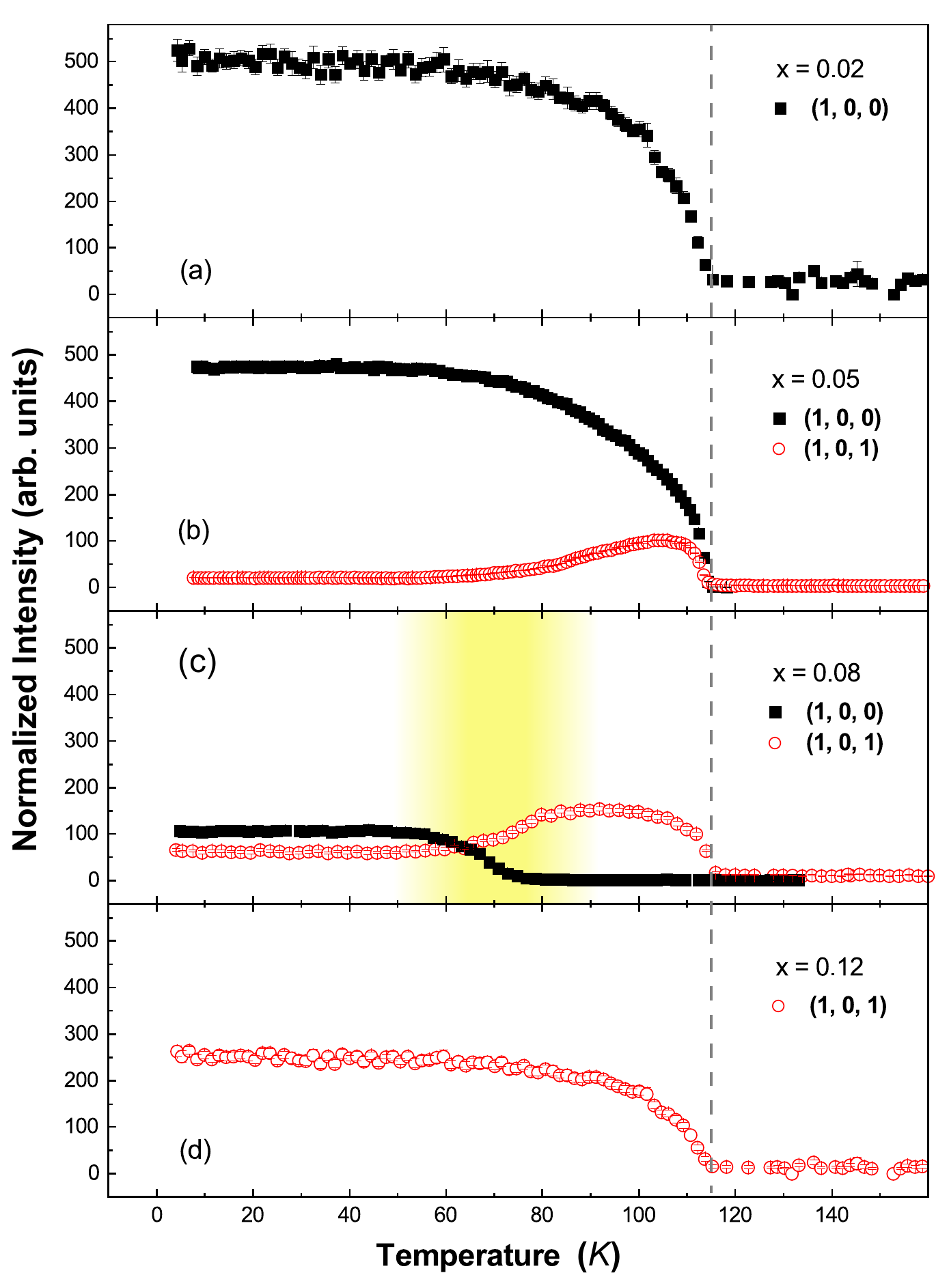}
\caption{(color online) Normalized intensities of magnetic Bragg peaks, plotted with the same scale, as a function of temperature. The $A$-centered phase, represented by the (1,0,0) reflection, is the only magnetic phase in the (a) x=0.02 crystal. The $B$-centered phase, represented by reflection (1,0,1), appears in (b) x=0.05 and coexists with (1,0,0). In x=0.08 (c), (1,0,1) continues to increase while (1,0,1) decreases. In x=0.12 (d), the $A$-centered phase is completely suppressed by the $B$-centered one. The vertical dashed lines shows the unchanged transition temperature. The shaded area in (c) shows the temperature range from the onset of $A$-phase and establishment of equilibrium between the two competing orders.
}
\end{figure}

\begin{table}
\caption{Magnetic properties of the Ca$_2$Ru$_{1-x}$Fe$_x$O$_4$ compounds.
}
\label{tab2}
\begin{ruledtabular}
\begin{tabular}{llccccc}
 x & x=0.02 & x=0.05  & x=0.08 &  x=0.12  \\

		 \hline
Spin &  & &  & \\				
Structure & $A$  & $A$ and $B$ & $A$ and $B$  & $B$\\
	
T$_N$(A) & 113 K & 113 K& 82 K & \\
T$_N$(B) &  & 113 K& 113 K & 113 K\\
m$_b$(A)   &1.62(7) &1.54(9) &0.73(4)& \\
m$_b$(B)   & &0.39(3) &0.71(4)& 1.38(7) \\
Summed &  & &  & \\				
Moment & 1.62(7)& 1.59(8) &1.02(8) &1.38(7) \\
\end{tabular}
\end{ruledtabular}
\end{table}

\begin{center}
{\bf D. Magnetic correlations in x=0.08}
\end{center}

The Fe-substitution releases stress from within the RuO$_6$ octahedra and reduces its distortion. Such structural change 
favors the $B$-type phase, where the nearest
next-layer neighbor, as indicated by the red line in Fig. 5, is ferromagnetically coupled. The nature of the rivalry between the two magnetic phases and their connection to the structural change can be better
elucidated by the detailed temperature dependence of 
magnetic correlation length. We chose the x=0.08 system for such effort because the phase battle it hosts is in full swing: they have comparable peak intensities and the waning $A$-phase has an altered transition temperature. Between 4 K and 150 K with a temperature step of 2 K we carried out $Q$-scans along the $K$- and $L$-directions across peak (0 1 1) and (0 1 2) which represent the $A$- and $B$-centered magnetic phases respectively. 
We find that the Gaussian line shape is the best fit for the scans across (0,1,2) and that Lorentzian for the scans across (0,1,1). The measured linewidth is the convolution of the scattering law and instrumental resolution function. None of the peaks has a width that is resolution limited, indicating finite magnetic correlations in both phases. Comparing with the x=0.02 system where these peaks all have Gaussian shape and their linewidths are all resolution-limited, we can conclude that the short-range correlation is caused by the doping induced magnetic competition. Extracting from the convolution with the instrument resolution, the $B$-phase correlation length $\xi/b$ from the $K$-scan or $\xi/c$ from the $L$-scan is given by $\sqrt{2ln(2)}/\pi\sigma$, where $\sigma$ is the value of the intrinsic Gaussian width.  
For the $A$-centered phase, an estimate of the correlation length $\xi$/b or $\xi$/c  of Lorentzian line-shape  is given by $2/\omega$, where $\omega$ is the Lorentzian width. The temperature dependence of the in-plane correlation length $\xi/b$ and out-of-plane length $\xi/c$ for the two magentic phases are displayed in Figure 9. 

\begin{figure}
\label{fig9}

\includegraphics [width=1.0\columnwidth]{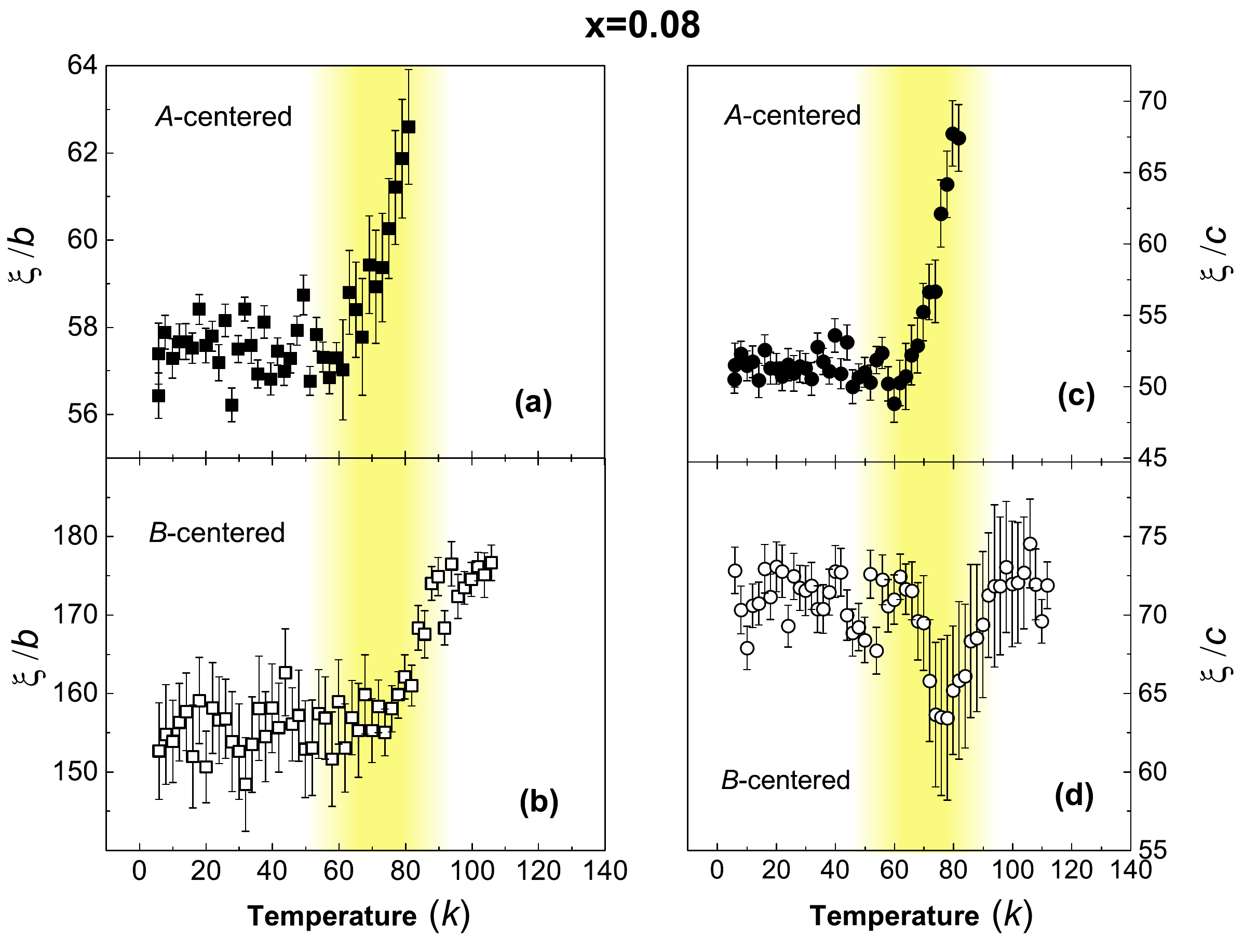}
\caption{(color online) The magnetic correlation length in the unit of cell numbers as a function of temperature in x=0.08 for (a) the $A$-centered phase along the $b$-axis, (b) the $B$-centered phase along the $b$-axis, (c) the the $A$-centered phase along the $c$-axis, and (d) the $B$-centered phase along the $c$-axis. The shaded areas show the same temperature range as in Fig. 8(c) where the two magnetic phases battle to reach an equilibrium.
}
\end{figure}

From Fig. 8(c) we know 80 K is about the transition temperature for phase-$A$ and 60 K is roughly
where the proportion of the two phases reaches equilibrium. So the shaded area in Fig. 9 shows where the two phases battle for intensity.
Both phases exhibit short-range magnetic correlation. Their competition further reduces their correlation lengths and prevents the development of long-range order. On cooling, the $B$-phase sets in at 113 K with in-plane correlation of about 180 unit cells, which remains constant until the onset of the $A$-phase at about 80 K, as shown in Fig. 9(b). $\xi/b$ in both phases decrease in the shaded temperature range before flatting out at lower temperature as the phase competition reaches equilibrium below 60 K. The out-of-plane correlation has similar temperature dependence except $\xi/c$ in the $B$-phase is not longer than that in the $A$-phase and recovers to its original value below 60 K after a dip caused by the competition in the shaded range as shown Fig.9(d).

\begin{figure}
\label{fig10}
\includegraphics [width=1.0\columnwidth]{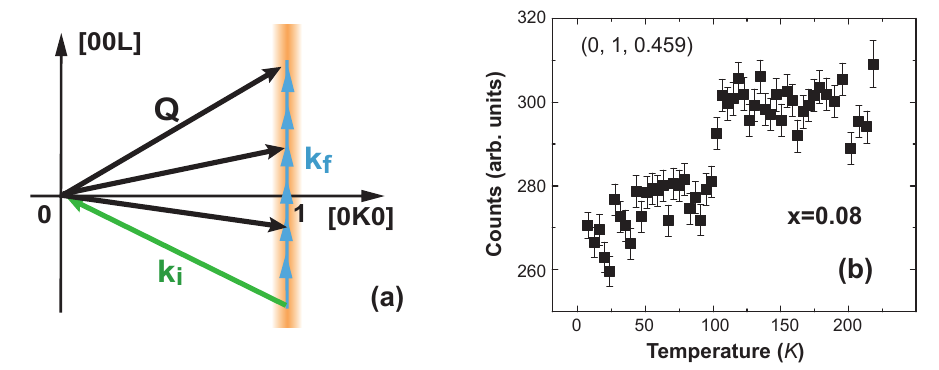}
\caption{(color online) (a) Momentum space diagram illustrating the two-axis energy integration method for a two-dimensional magnetic system. (b) The energy-integrated intensities as a function of temperature collected using the two-axis mode on HB3 shows the absence of the 2D fluctuation above $T_N$.
}
\end{figure}

Although the magnetic correlation extends in three dimensions, the interlayer coupling in the parent compound is only 0.03 meV, much weaker than the coupling within the planes which is 8 meV for the nearest-neighbor exchange. \cite {Kunkemoller15} Perturbed Angular Correlation measurement also suggested two-dimensional character of the magnetic ordering in the parent compound. \cite{Rams} Spin ordering in similar structures, such as high-T$_c$ cuprate La$_2$CuO$_4$ \cite{Endoh} and iron-pnictide BaFe$_2$As$_2$ \cite{Wilson}, have been demonstrated to arise from two-dimensional spin fluctuations. In the Fe-dope CRO compound, the two phases further 
undermines each other resulting in two separate transitions. Critical fluctuations in proximity to the onset of either phase can provide important insight into the nature of the magnetism in CRO. We explored possibility of 2D fluctuations using the two-axis energy integration method as illustrated in Figure 10 (a). The HB3 triple axis spectrometer is put into the two-axis energy-integrated mode by removing the analyzer, lining up the detector with analyzer arm so that the energy transfers probed by neutron energy loss are integrated up $E_i$ which is 14.7 meV in this case. A PG filter was used before the sample 
to remove higher-harmonic neutrons. If the 2D order exists in the paramagnetic regime, it would appear in the momentum space as a ridge that extends perpendicular to $(H,K,0)$ plane.\cite{Birgeneau} 
In the (0, $K,L$) scattering plane, a scan along $K$ above $T_N$ would reveal a peak if the 2D scattering ridge does exist. To ensure the detector, aligned along $\textbf{\textit{k}}_f$, is parallel to $L$ at each point of the scan, $L$ needs to satisfy $L=\frac{c}{\lambda}-c\sqrt{\frac{1}{\lambda^2}-\frac{K^2}{b^2}}$, where $b$, $c$ are lattice constants and $\lambda$ the wavelength of the incident neutron beam. 
As the system is cooled toward $T_N$, the ridge should decrease in length but grow in the scattering intensity. A sharp drop of intensity is expected across $T_N$ as the critical scattering condenses in to the 3D magnetic Bragg positions. However, our $K$-scans across (0,1,0.4594) at 120 K, 130 K and 150 K show no peak. Neither does the temperature dependence of energy integrated intensity at (0,1,0.4594), as shown in Fig. 9(b). The abrupt increase of intensity around 113 K should be attributed to diffuse paramagnetic scattering above the magnetic transition temperature. \cite{Wollan} The absence of the scattering ridge  suggest that the competing magnetic phases neither cause nor arise from two-dimensional spin fluctuations. 

\begin{center}
{\bf {IV}. DISCUSSION}
\end{center}

The absence of the 2D AFM critical fluctuation near the magnetic transition temperature in the x=0.08 system rules out the two magnetic phases as merely variations on the stacking sequence of the individual AFM RuO$_2$ layers. The development of magnetic correlation below T$_N$ also exhibits three-dimensional (3D) character even before the equilibrium of the two phase is reached below 60 K. The magnetic coexistence of two phases in x=0.05 and 0.08 should be phase separation in nature where inhomogeneous octahedral distortions promote different magnetic instability and form 3D magnetic domains. We note that $T_N$ in the $A$-centered phase of x=0.02 and in $B$-centered phase of x=0.12 remains the same, implying the unaffected exchange energy.  In the Van Vleck-type excitonic magnetism, \cite{Akbari} the condensate intensities and staggered moment essentially depends on the $J/ \Lambda$ ratio, where $J$ is the exchange energy scale and $\Lambda$ the SOC parameter. The reduced $\Lambda$ by adding $3d$ Fe is expected to cause an increased ordered moment. The observed decrease should be attributed to the disorder caused by the phase competition. Such disorder is also evidenced by the short-range magnetic correlations in x=0.08 as well as the recovery of moment in x=0.12 as the phase competition ends with the complete dominance of the $B$-centered phase.  The 3D short-range correlation can hardly be justified by the weak interplanar exchange coupling \cite{Kunkemoller15} within the framework of the Heisenberg model. Neither can spins alone account for the anomalous release of the apical compression, $\delta$, by the onset of magnetic order in x=0.08 [Fig. 4(a)]. This anomalous lattice response suggests the important role of the orbital polarization which can be controlled by structure. \cite{Fang, Anisimov} To understand how the Fe-induced octahedral relaxation could eventually lead to the dominance of the $B$-centered spin order, we discuss below the effect of these structural distortions individually.

The compressed RuO$_6$ is the Jahn-Teller distortion that lifts the $t_{2g}$ degeneracy by lowering the $xy$ orbitals relative to the $yz$ and $zx$ orbitals. Consequently, the $xy$ orbitals are fully populated, leading to the insulating state. Any elongation of the $c$-axis induced by uni-axial pressure \cite{Taniguchi} or chemical doping \cite{Friedt, Carlo, Nakatsuji2000, Pincini18, Pincini19} leads a transition to a metallic state in which the $xy$ orbitals are only partially occupied. The octahedral flattening also stabilizes magnetic order, both FM and AFM, by increasing DOS at the Fermi level for the former and shifting nesting vector through orbital polarization for the latter. \cite{Fang} The RuO$_6$ octahedra remain flattened up to 12$\%$ Fe substitution, thus explaining the persistence of the magnetic orders. An inversion of flattened distortion changes the sign of the crystal field potential and results in the modification of charge distribution. Such charge transfer could directly change the preferential occupation in the orbitals of the $t_{2g}$ sector, facilitating a switch to a different orbital order and a new spin structure. \cite{Cuoco1}The ratio of the apical to planar bond length, $\delta$, in the parent CRO compound is already in the proximity of the critical value for the transition from the AFM to FM instabilities as evidenced by the FM order induced by pressure or magnetic field.\cite{Cao97, Nakamura, Steffens05} 
With further increased $\delta$ by Fe-doping, charge transfer may occur from the doubly occupied $xy$ sector to the $z$ sector of the $t_{2g}$ manifold as suggested by the abrupt change of $\delta$  in x=0.08. This reverse of $\delta$, for that matter, is observed at 100 K in the $B$-centered phase, but not in the parent CRO compound where the $A$-centered order dominates.

The apical elongation of the octahedron is not the necessary condition for the $A$ to $B$ phase transition though. For example, hydraulic pressure enhances the octahedral flattening in the parent CRO compound as it is driven into the $B$-centered phase. \cite{Steffens05} If the Fe-induced partial release of the compressive distortion is insufficient to trigger the transition of the orbital state, the spins are then fixed in the checker-board like AFM pattern in the RuO$_2$ plane, leaving the inter-planar exchange coupling to the mercy of small perturbations. Such perturbation can be provided by the released octahedral rotation and tilt. 

The rotation $\phi$ undermines the hybridization between the O(1)-2$p$ and the $d_{xy}$ state, lowers and narrows the $d_{xy}$ band. \cite{Fang} The resultant increase of DOS at Fermi level facilitates FM instability. The tilt $\theta$, on the other hand, narrows all $t_{2g}$ bands, \cite{Mizokawa} therefore enhances nesting so promotes AFM instability.\cite{Fang} The combination of $\phi$ and $\theta$ is responsible for the enhancement of the AFM instability. The release of the strapped octahedron by Fe-substitution, reflected by the reduced $\phi$ and $\theta$, should lead to the suppressed AFM order and enhanced FM fluctuation. Consequently, the nearest neighbor exchange interaction, highlighted by the red line in Fig. 5, becomes FM. The reduction of AFM order is supported by the overall reduction of the ordered moment, the decreased T$_N$ of the $A$-phase, and eventually decreased T$_N$ of the $B$-phase in x=0.2. \cite{Yuan}  Our finding fits the trend of such released octahedral distortion realized either by doping Ca with bigger ions, such as Sr \cite{Friedt, Carlo, Nakatsuji2000}  and La \cite{Pincini18, Pincini19}, or by hydraulic pressure. \cite{Steffens05} Although the increased Ru-O(1)-Ru bond angle as a result of octahedral release is supposed to enhance the planar AFM superexchange constant, the increase is too small to reverse the trend.

The octahedral rotation and tilt could cause the canting of the Ru moment through the Dzyaloshinsky-Moriya interaction. 
With SOC included in the spin Hamiltonian, the exchange interaction has a antisymmetric term that contains the cross product of two neighboring spins. Energy can be gained by having a finite angle between the two spins. For the parent CRO, the canting along the $a$- \cite{Braden1} and the $c$-axis \cite{Fukazawa} have been proposed to explain the magnetic susceptibility data. The direct observation of the $c$-axis canting was reported by a recent resonant elastic x-ray scattering study.  \cite{Porter}
Although the refined direction of the ordered moment in this investigation is along the $b$-axis, a tiny canting component along the other axes could not be completely ruled out due to the limited number of available magnetic peaks.  The canting moments should cancel out antiferromagnetically between the RuO$_2$ layers in the $A$-centered phase, while add up to finite FM component in the $B$-centered one. As Fe-doping increases, the reduced rotation and tilt should further reduce the canting angle, if it exist at all, making it even more difficult to detect. In the scope of the current study, the role of SOC in driving the magnetic phase from $A$ to $B$ is inconsequential because the canting is allowed in both magnetic space group and would not favor one phase over the other. Ultimately the FM instability needs to be promoted to have the $B$-centered phase and that, as argued earlier, is provided by the reduced rotation and tilt with or without the moment canting.

\begin{center}
{\bf {V}. CONCLUSIONS}
\end{center}
 
In conclusion, the diffraction measurements clearly establish the crystal and the spin structures of four CRO compounds with various Fe-dopings. As Fe-substitution gradually relaxes the octahedral distortion, the $A$-centered magnetic phase concedes to the emergent $B$-centered one which eventually dominates as the Fe-doping reaches 0.12. The ordered moment of the two phases, either in dominance or in coexistence, all lies along the $b$-axis. In x=0.08 where the two magnetic phases are in close division, the intra- and inter-planar spin correlations become short-ranged and the total moment get reduced. The critical scattering measurement did not detect any 2D spin fluctuation above $T_N$. Also in this doping we observed an abrupt partial release of octahedral flattening across $T_N$. The character of the magnetic competition and the structural response implies the essential role of spin orbital correlation in determining the magnetic ground state. 

\begin{center}
{\bf  ACKNOWLEDGEMENT}
\end{center}

This research used resources at the High Flux Isotope Reactor, a DOE Office of Science User Facility operated by the Oak Ridge National Laboratory, and was supported in part by the US Department of Energy, Office of Science, Office of Basic Energy Sciences. GC acknowledges support by the National Science Foundation via grant DMR-1903888.

\bibliography{Ca2RuFeO4}

\begin{thebibliography}{60}
\expandafter\ifx\csname natexlab\endcsname\relax\def\natexlab#1{#1}\fi
\expandafter\ifx\csname bibnamefont\endcsname\relax
  \def\bibnamefont#1{#1}\fi
\expandafter\ifx\csname bibfnamefont\endcsname\relax
  \def\bibfnamefont#1{#1}\fi
\expandafter\ifx\csname citenamefont\endcsname\relax
  \def\citenamefont#1{#1}\fi
\expandafter\ifx\csname url\endcsname\relax
  \def\url#1{\texttt{#1}}\fi
\expandafter\ifx\csname urlprefix\endcsname\relax\def\urlprefix{URL }\fi
\providecommand{\bibinfo}[2]{#2}
\providecommand{\eprint}[2][]{\url{#2}}

\bibitem[{\citenamefont{Nakatsuji et~al.}(1997)\citenamefont{Nakatsuji, Ikeda,
  and Maeno}}]{Nakatsuji97}
\bibinfo{author}{\bibfnamefont{S.}~\bibnamefont{Nakatsuji}},
  \bibinfo{author}{\bibfnamefont{S.~I.} \bibnamefont{Ikeda}}, \bibnamefont{and}
  \bibinfo{author}{\bibfnamefont{Y.}~\bibnamefont{Maeno}},
  \bibinfo{journal}{Journal of the Physical Society of Japan}
  \textbf{\bibinfo{volume}{66}}, \bibinfo{pages}{1868} (\bibinfo{year}{1997}).

\bibitem[{\citenamefont{Gorelov et~al.}(2010)\citenamefont{Gorelov, Karolak,
  Wehling, Lechermann, Lichtenstein, and Pavarini}}]{Gorelov}
\bibinfo{author}{\bibfnamefont{E.}~\bibnamefont{Gorelov}},
  \bibinfo{author}{\bibfnamefont{M.}~\bibnamefont{Karolak}},
  \bibinfo{author}{\bibfnamefont{T.~O.} \bibnamefont{Wehling}},
  \bibinfo{author}{\bibfnamefont{F.}~\bibnamefont{Lechermann}},
  \bibinfo{author}{\bibfnamefont{A.~I.} \bibnamefont{Lichtenstein}},
  \bibnamefont{and} \bibinfo{author}{\bibfnamefont{E.}~\bibnamefont{Pavarini}},
  \bibinfo{journal}{Physical Review Letters} \textbf{\bibinfo{volume}{104}},
  \bibinfo{pages}{226401} (\bibinfo{year}{2010}).

\bibitem[{\citenamefont{Liebsch and Ishida}(2007)}]{Liebsch}
\bibinfo{author}{\bibfnamefont{A.}~\bibnamefont{Liebsch}} \bibnamefont{and}
  \bibinfo{author}{\bibfnamefont{H.}~\bibnamefont{Ishida}},
  \bibinfo{journal}{Physical Review Letters} \textbf{\bibinfo{volume}{98}},
  \bibinfo{pages}{216403} (\bibinfo{year}{2007}).

\bibitem[{\citenamefont{Hotta and Dagotto}(2001)}]{Hotta}
\bibinfo{author}{\bibfnamefont{T.}~\bibnamefont{Hotta}} \bibnamefont{and}
  \bibinfo{author}{\bibfnamefont{E.}~\bibnamefont{Dagotto}},
  \bibinfo{journal}{Physical Review Letters} \textbf{\bibinfo{volume}{88}},
  \bibinfo{pages}{017201} (\bibinfo{year}{2001}).

\bibitem[{\citenamefont{Sutter et~al.}(2017)\citenamefont{Sutter, Fatuzzo,
  Moser, Kim, Fittipaldi, Vecchione, Granata, Sassa, Cossalter, Gatti
  et~al.}}]{Sutter}
\bibinfo{author}{\bibfnamefont{D.}~\bibnamefont{Sutter}},
  \bibinfo{author}{\bibfnamefont{C.~G.} \bibnamefont{Fatuzzo}},
  \bibinfo{author}{\bibfnamefont{S.}~\bibnamefont{Moser}},
  \bibinfo{author}{\bibfnamefont{M.}~\bibnamefont{Kim}},
  \bibinfo{author}{\bibfnamefont{R.}~\bibnamefont{Fittipaldi}},
  \bibinfo{author}{\bibfnamefont{A.}~\bibnamefont{Vecchione}},
  \bibinfo{author}{\bibfnamefont{V.}~\bibnamefont{Granata}},
  \bibinfo{author}{\bibfnamefont{Y.}~\bibnamefont{Sassa}},
  \bibinfo{author}{\bibfnamefont{F.}~\bibnamefont{Cossalter}},
  \bibinfo{author}{\bibfnamefont{G.}~\bibnamefont{Gatti}},
  \bibnamefont{et~al.}, \bibinfo{journal}{Nature Communications}
  \textbf{\bibinfo{volume}{8}}, \bibinfo{pages}{15176} (\bibinfo{year}{2017}).

\bibitem[{\citenamefont{Anisimov et~al.}(2002)\citenamefont{Anisimov, Nekrasov,
  Kondakov, Rice, and Sigrist}}]{Anisimov}
\bibinfo{author}{\bibfnamefont{V.~I.} \bibnamefont{Anisimov}},
  \bibinfo{author}{\bibfnamefont{I.~A.} \bibnamefont{Nekrasov}},
  \bibinfo{author}{\bibfnamefont{D.~E.} \bibnamefont{Kondakov}},
  \bibinfo{author}{\bibfnamefont{T.~M.} \bibnamefont{Rice}}, \bibnamefont{and}
  \bibinfo{author}{\bibfnamefont{M.}~\bibnamefont{Sigrist}},
  \bibinfo{journal}{European Physical Journal B} \textbf{\bibinfo{volume}{25}},
  \bibinfo{pages}{191} (\bibinfo{year}{2002}).

\bibitem[{\citenamefont{Neupane et~al.}(2009)\citenamefont{Neupane, Richard,
  Pan, Xu, Jin, Mandrus, Dai, Fang, Wang, and Ding}}]{Neupane}
\bibinfo{author}{\bibfnamefont{M.}~\bibnamefont{Neupane}},
  \bibinfo{author}{\bibfnamefont{P.}~\bibnamefont{Richard}},
  \bibinfo{author}{\bibfnamefont{Z.~H.} \bibnamefont{Pan}},
  \bibinfo{author}{\bibfnamefont{Y.~M.} \bibnamefont{Xu}},
  \bibinfo{author}{\bibfnamefont{R.}~\bibnamefont{Jin}},
  \bibinfo{author}{\bibfnamefont{D.}~\bibnamefont{Mandrus}},
  \bibinfo{author}{\bibfnamefont{X.}~\bibnamefont{Dai}},
  \bibinfo{author}{\bibfnamefont{Z.}~\bibnamefont{Fang}},
  \bibinfo{author}{\bibfnamefont{Z.}~\bibnamefont{Wang}}, \bibnamefont{and}
  \bibinfo{author}{\bibfnamefont{H.}~\bibnamefont{Ding}},
  \bibinfo{journal}{Physical Review Letters} \textbf{\bibinfo{volume}{103}},
  \bibinfo{pages}{097001} (\bibinfo{year}{2009}).

\bibitem[{\citenamefont{Pavarini et~al.}(2004)\citenamefont{Pavarini, Biermann,
  Poteryaev, Lichtenstein, Georges, and Andersen}}]{Pavarini04}
\bibinfo{author}{\bibfnamefont{E.}~\bibnamefont{Pavarini}},
  \bibinfo{author}{\bibfnamefont{S.}~\bibnamefont{Biermann}},
  \bibinfo{author}{\bibfnamefont{A.}~\bibnamefont{Poteryaev}},
  \bibinfo{author}{\bibfnamefont{A.~I.} \bibnamefont{Lichtenstein}},
  \bibinfo{author}{\bibfnamefont{A.}~\bibnamefont{Georges}}, \bibnamefont{and}
  \bibinfo{author}{\bibfnamefont{O.~K.} \bibnamefont{Andersen}},
  \bibinfo{journal}{Physical Review Letters} \textbf{\bibinfo{volume}{92}},
  \bibinfo{pages}{176403} (\bibinfo{year}{2004}).

\bibitem[{\citenamefont{Haule and Kotliar}(2009)}]{Haule}
\bibinfo{author}{\bibfnamefont{K.}~\bibnamefont{Haule}} \bibnamefont{and}
  \bibinfo{author}{\bibfnamefont{G.}~\bibnamefont{Kotliar}},
  \bibinfo{journal}{New Journal of Physics} \textbf{\bibinfo{volume}{11}},
  \bibinfo{pages}{025021} (\bibinfo{year}{2009}).

\bibitem[{\citenamefont{Werner and Millis}(2007)}]{Werner}
\bibinfo{author}{\bibfnamefont{P.}~\bibnamefont{Werner}} \bibnamefont{and}
  \bibinfo{author}{\bibfnamefont{A.~J.} \bibnamefont{Millis}},
  \bibinfo{journal}{Physical Review Letters} \textbf{\bibinfo{volume}{99}},
  \bibinfo{pages}{126405} (\bibinfo{year}{2007}).

\bibitem[{\citenamefont{de' Medici}(2011)}]{Medici}
\bibinfo{author}{\bibfnamefont{L.}~\bibnamefont{de' Medici}},
  \bibinfo{journal}{Physical Review B} \textbf{\bibinfo{volume}{83}},
  \bibinfo{pages}{205112} (\bibinfo{year}{2011}).

\bibitem[{\citenamefont{Chi et~al.}(2016)\citenamefont{Chi, Uwatoko, Cao,
  Hirata, Hashizume, Aoyama, and Ohgushi}}]{Chi}
\bibinfo{author}{\bibfnamefont{S.}~\bibnamefont{Chi}},
  \bibinfo{author}{\bibfnamefont{Y.}~\bibnamefont{Uwatoko}},
  \bibinfo{author}{\bibfnamefont{H.~B.} \bibnamefont{Cao}},
  \bibinfo{author}{\bibfnamefont{Y.}~\bibnamefont{Hirata}},
  \bibinfo{author}{\bibfnamefont{K.}~\bibnamefont{Hashizume}},
  \bibinfo{author}{\bibfnamefont{T.}~\bibnamefont{Aoyama}}, \bibnamefont{and}
  \bibinfo{author}{\bibfnamefont{K.}~\bibnamefont{Ohgushi}},
  \bibinfo{journal}{Physical Review Letters} \textbf{\bibinfo{volume}{117}},
  \bibinfo{pages}{047003} (\bibinfo{year}{2016}).

\bibitem[{\citenamefont{Liu}(2011)}]{Liu11}
\bibinfo{author}{\bibfnamefont{G.-Q.} \bibnamefont{Liu}},
  \bibinfo{journal}{Physical Review B} \textbf{\bibinfo{volume}{84}},
  \bibinfo{pages}{235136} (\bibinfo{year}{2011}).

\bibitem[{\citenamefont{Mizokawa et~al.}(2001)\citenamefont{Mizokawa, Tjeng,
  Sawatzky, Ghiringhelli, Tjernberg, Brookes, Fukazawa, Nakatsuji, and
  Maeno}}]{Mizokawa}
\bibinfo{author}{\bibfnamefont{T.}~\bibnamefont{Mizokawa}},
  \bibinfo{author}{\bibfnamefont{L.~H.} \bibnamefont{Tjeng}},
  \bibinfo{author}{\bibfnamefont{G.~A.} \bibnamefont{Sawatzky}},
  \bibinfo{author}{\bibfnamefont{G.}~\bibnamefont{Ghiringhelli}},
  \bibinfo{author}{\bibfnamefont{O.}~\bibnamefont{Tjernberg}},
  \bibinfo{author}{\bibfnamefont{N.~B.} \bibnamefont{Brookes}},
  \bibinfo{author}{\bibfnamefont{H.}~\bibnamefont{Fukazawa}},
  \bibinfo{author}{\bibfnamefont{S.}~\bibnamefont{Nakatsuji}},
  \bibnamefont{and} \bibinfo{author}{\bibfnamefont{Y.}~\bibnamefont{Maeno}},
  \bibinfo{journal}{Physical Review Letters} \textbf{\bibinfo{volume}{87}},
  \bibinfo{pages}{077202} (\bibinfo{year}{2001}).

\bibitem[{\citenamefont{Witczak-Krempa
  et~al.}(2014)\citenamefont{Witczak-Krempa, Chen, Kim, and Balents}}]{Witczak}
\bibinfo{author}{\bibfnamefont{W.}~\bibnamefont{Witczak-Krempa}},
  \bibinfo{author}{\bibfnamefont{G.}~\bibnamefont{Chen}},
  \bibinfo{author}{\bibfnamefont{Y.~B.} \bibnamefont{Kim}}, \bibnamefont{and}
  \bibinfo{author}{\bibfnamefont{L.}~\bibnamefont{Balents}},
  \bibinfo{journal}{Annual Review of Condensed Matter Physics, Vol 5}
  \textbf{\bibinfo{volume}{5}}, \bibinfo{pages}{57} (\bibinfo{year}{2014}).

\bibitem[{\citenamefont{Cao et~al.}(1997)\citenamefont{Cao, McCall, Shepard,
  Crow, and Guertin}}]{Cao97}
\bibinfo{author}{\bibfnamefont{G.}~\bibnamefont{Cao}},
  \bibinfo{author}{\bibfnamefont{S.}~\bibnamefont{McCall}},
  \bibinfo{author}{\bibfnamefont{M.}~\bibnamefont{Shepard}},
  \bibinfo{author}{\bibfnamefont{J.~E.} \bibnamefont{Crow}}, \bibnamefont{and}
  \bibinfo{author}{\bibfnamefont{R.~P.} \bibnamefont{Guertin}},
  \bibinfo{journal}{Physical Review B} \textbf{\bibinfo{volume}{56}},
  \bibinfo{pages}{R2916} (\bibinfo{year}{1997}).

\bibitem[{\citenamefont{Alexander et~al.}(1999)\citenamefont{Alexander, Cao,
  Dobrosavljevic, McCall, Crow, Lochner, and Guertin}}]{Alexander}
\bibinfo{author}{\bibfnamefont{C.~S.} \bibnamefont{Alexander}},
  \bibinfo{author}{\bibfnamefont{G.}~\bibnamefont{Cao}},
  \bibinfo{author}{\bibfnamefont{V.}~\bibnamefont{Dobrosavljevic}},
  \bibinfo{author}{\bibfnamefont{S.}~\bibnamefont{McCall}},
  \bibinfo{author}{\bibfnamefont{J.~E.} \bibnamefont{Crow}},
  \bibinfo{author}{\bibfnamefont{E.}~\bibnamefont{Lochner}}, \bibnamefont{and}
  \bibinfo{author}{\bibfnamefont{R.~P.} \bibnamefont{Guertin}},
  \bibinfo{journal}{Physical Review B} \textbf{\bibinfo{volume}{60}},
  \bibinfo{pages}{R8422} (\bibinfo{year}{1999}).

\bibitem[{\citenamefont{Braden et~al.}(1998)\citenamefont{Braden, Andre,
  Nakatsuji, and Maeno}}]{Braden1}
\bibinfo{author}{\bibfnamefont{M.}~\bibnamefont{Braden}},
  \bibinfo{author}{\bibfnamefont{G.}~\bibnamefont{Andre}},
  \bibinfo{author}{\bibfnamefont{S.}~\bibnamefont{Nakatsuji}},
  \bibnamefont{and} \bibinfo{author}{\bibfnamefont{Y.}~\bibnamefont{Maeno}},
  \bibinfo{journal}{Physical Review B} \textbf{\bibinfo{volume}{58}},
  \bibinfo{pages}{847} (\bibinfo{year}{1998}).

\bibitem[{\citenamefont{Nakatsuji and Maeno}(2000)}]{Nakatsuji2000}
\bibinfo{author}{\bibfnamefont{S.}~\bibnamefont{Nakatsuji}} \bibnamefont{and}
  \bibinfo{author}{\bibfnamefont{Y.}~\bibnamefont{Maeno}},
  \bibinfo{journal}{Physical Review B} \textbf{\bibinfo{volume}{62}},
  \bibinfo{pages}{6458} (\bibinfo{year}{2000}), ISSN \bibinfo{issn}{1098-0121}.

\bibitem[{\citenamefont{Fatuzzo et~al.}(2015)\citenamefont{Fatuzzo, Dantz,
  Fatale, Olalde-Velasco, Shaik, Dalla~Piazza, Toth, Pelliciari, Fittipaldi,
  Vecchione et~al.}}]{Fatuzzo}
\bibinfo{author}{\bibfnamefont{C.~G.} \bibnamefont{Fatuzzo}},
  \bibinfo{author}{\bibfnamefont{M.}~\bibnamefont{Dantz}},
  \bibinfo{author}{\bibfnamefont{S.}~\bibnamefont{Fatale}},
  \bibinfo{author}{\bibfnamefont{P.}~\bibnamefont{Olalde-Velasco}},
  \bibinfo{author}{\bibfnamefont{N.~E.} \bibnamefont{Shaik}},
  \bibinfo{author}{\bibfnamefont{B.}~\bibnamefont{Dalla~Piazza}},
  \bibinfo{author}{\bibfnamefont{S.}~\bibnamefont{Toth}},
  \bibinfo{author}{\bibfnamefont{J.}~\bibnamefont{Pelliciari}},
  \bibinfo{author}{\bibfnamefont{R.}~\bibnamefont{Fittipaldi}},
  \bibinfo{author}{\bibfnamefont{A.}~\bibnamefont{Vecchione}},
  \bibnamefont{et~al.}, \bibinfo{journal}{Physical Review B}
  \textbf{\bibinfo{volume}{91}}, \bibinfo{pages}{155104}
  (\bibinfo{year}{2015}).

\bibitem[{\citenamefont{Khaliullin}(2013)}]{Khaliullin}
\bibinfo{author}{\bibfnamefont{G.}~\bibnamefont{Khaliullin}},
  \bibinfo{journal}{Physical Review Letters} \textbf{\bibinfo{volume}{111}},
  \bibinfo{pages}{197201} (\bibinfo{year}{2013}).

\bibitem[{\citenamefont{Kunkemoller et~al.}(2015)\citenamefont{Kunkemoller,
  Khomskii, Steffens, Piovano, Nugroho, and Braden}}]{Kunkemoller15}
\bibinfo{author}{\bibfnamefont{S.}~\bibnamefont{Kunkemoller}},
  \bibinfo{author}{\bibfnamefont{D.}~\bibnamefont{Khomskii}},
  \bibinfo{author}{\bibfnamefont{P.}~\bibnamefont{Steffens}},
  \bibinfo{author}{\bibfnamefont{A.}~\bibnamefont{Piovano}},
  \bibinfo{author}{\bibfnamefont{A.~A.} \bibnamefont{Nugroho}},
  \bibnamefont{and} \bibinfo{author}{\bibfnamefont{M.}~\bibnamefont{Braden}},
  \bibinfo{journal}{Physical Review Letters} \textbf{\bibinfo{volume}{115}},
  \bibinfo{pages}{247201} (\bibinfo{year}{2015}).

\bibitem[{\citenamefont{Kunkemoller et~al.}(2017)\citenamefont{Kunkemoller,
  Komleva, Streltsov, Hoffmann, Khomskii, Steffens, Sidis, Schmalzl, and
  Braden}}]{Kunkemoller17}
\bibinfo{author}{\bibfnamefont{S.}~\bibnamefont{Kunkemoller}},
  \bibinfo{author}{\bibfnamefont{E.}~\bibnamefont{Komleva}},
  \bibinfo{author}{\bibfnamefont{S.~V.} \bibnamefont{Streltsov}},
  \bibinfo{author}{\bibfnamefont{S.}~\bibnamefont{Hoffmann}},
  \bibinfo{author}{\bibfnamefont{D.~I.} \bibnamefont{Khomskii}},
  \bibinfo{author}{\bibfnamefont{P.}~\bibnamefont{Steffens}},
  \bibinfo{author}{\bibfnamefont{Y.}~\bibnamefont{Sidis}},
  \bibinfo{author}{\bibfnamefont{K.}~\bibnamefont{Schmalzl}}, \bibnamefont{and}
  \bibinfo{author}{\bibfnamefont{M.}~\bibnamefont{Braden}},
  \bibinfo{journal}{Physical Review B} \textbf{\bibinfo{volume}{95}},
  \bibinfo{pages}{214408} (\bibinfo{year}{2017}).

\bibitem[{\citenamefont{Jain et~al.}(2017)\citenamefont{Jain, Krautloher,
  Porras, Ryu, Chen, Abernathy, Park, Ivanov, Chaloupka, Khallin
  et~al.}}]{Jain}
\bibinfo{author}{\bibfnamefont{A.}~\bibnamefont{Jain}},
  \bibinfo{author}{\bibfnamefont{M.}~\bibnamefont{Krautloher}},
  \bibinfo{author}{\bibfnamefont{J.}~\bibnamefont{Porras}},
  \bibinfo{author}{\bibfnamefont{G.~H.} \bibnamefont{Ryu}},
  \bibinfo{author}{\bibfnamefont{D.~P.} \bibnamefont{Chen}},
  \bibinfo{author}{\bibfnamefont{D.~L.} \bibnamefont{Abernathy}},
  \bibinfo{author}{\bibfnamefont{J.~T.} \bibnamefont{Park}},
  \bibinfo{author}{\bibfnamefont{A.}~\bibnamefont{Ivanov}},
  \bibinfo{author}{\bibfnamefont{J.}~\bibnamefont{Chaloupka}},
  \bibinfo{author}{\bibfnamefont{G.}~\bibnamefont{Khallin}},
  \bibnamefont{et~al.}, \bibinfo{journal}{Nature Physics}
  \textbf{\bibinfo{volume}{13}}, \bibinfo{pages}{633} (\bibinfo{year}{2017}).

\bibitem[{\citenamefont{Zhang and Pavarini}(2017)}]{Zhang17}
\bibinfo{author}{\bibfnamefont{G.~R.} \bibnamefont{Zhang}} \bibnamefont{and}
  \bibinfo{author}{\bibfnamefont{E.}~\bibnamefont{Pavarini}},
  \bibinfo{journal}{Physical Review B} \textbf{\bibinfo{volume}{95}},
  \bibinfo{pages}{075145} (\bibinfo{year}{2017}).

\bibitem[{\citenamefont{Akbari and Khaliullin}(2014)}]{Akbari}
\bibinfo{author}{\bibfnamefont{A.}~\bibnamefont{Akbari}} \bibnamefont{and}
  \bibinfo{author}{\bibfnamefont{G.}~\bibnamefont{Khaliullin}},
  \bibinfo{journal}{Physical Review B} \textbf{\bibinfo{volume}{90}},
  \bibinfo{pages}{035137} (\bibinfo{year}{2014}).

\bibitem[{\citenamefont{Souliou et~al.}(2017)\citenamefont{Souliou, Chaloupka,
  Khaliullin, Ryu, Jain, Kim, Le~Tacon, and Keimer}}]{Souliou}
\bibinfo{author}{\bibfnamefont{S.~M.} \bibnamefont{Souliou}},
  \bibinfo{author}{\bibfnamefont{J.}~\bibnamefont{Chaloupka}},
  \bibinfo{author}{\bibfnamefont{G.}~\bibnamefont{Khaliullin}},
  \bibinfo{author}{\bibfnamefont{G.}~\bibnamefont{Ryu}},
  \bibinfo{author}{\bibfnamefont{A.}~\bibnamefont{Jain}},
  \bibinfo{author}{\bibfnamefont{B.~J.} \bibnamefont{Kim}},
  \bibinfo{author}{\bibfnamefont{M.}~\bibnamefont{Le~Tacon}}, \bibnamefont{and}
  \bibinfo{author}{\bibfnamefont{B.}~\bibnamefont{Keimer}},
  \bibinfo{journal}{Physical Review Letters} \textbf{\bibinfo{volume}{119}},
  \bibinfo{pages}{067201} (\bibinfo{year}{2017}), ISSN
  \bibinfo{issn}{0031-9007}.

\bibitem[{\citenamefont{Das et~al.}(2018)\citenamefont{Das, Forte, Fittipaldi,
  Fatuzzo, Granata, Ivashko, Horio, Schindler, Dantz, Tseng et~al.}}]{Das}
\bibinfo{author}{\bibfnamefont{L.}~\bibnamefont{Das}},
  \bibinfo{author}{\bibfnamefont{F.}~\bibnamefont{Forte}},
  \bibinfo{author}{\bibfnamefont{R.}~\bibnamefont{Fittipaldi}},
  \bibinfo{author}{\bibfnamefont{C.~G.} \bibnamefont{Fatuzzo}},
  \bibinfo{author}{\bibfnamefont{V.}~\bibnamefont{Granata}},
  \bibinfo{author}{\bibfnamefont{O.}~\bibnamefont{Ivashko}},
  \bibinfo{author}{\bibfnamefont{M.}~\bibnamefont{Horio}},
  \bibinfo{author}{\bibfnamefont{F.}~\bibnamefont{Schindler}},
  \bibinfo{author}{\bibfnamefont{M.}~\bibnamefont{Dantz}},
  \bibinfo{author}{\bibfnamefont{Y.}~\bibnamefont{Tseng}},
  \bibnamefont{et~al.}, \bibinfo{journal}{Physical Review X}
  \textbf{\bibinfo{volume}{8}}, \bibinfo{pages}{011048} (\bibinfo{year}{2018}).

\bibitem[{\citenamefont{Steffens et~al.}(2005)\citenamefont{Steffens, Friedt,
  Alireza, Marshall, Schmidt, Nakamura, Nakatsuji, Maeno, Lengsdorf,
  Abd-Elmeguid et~al.}}]{Steffens05}
\bibinfo{author}{\bibfnamefont{P.}~\bibnamefont{Steffens}},
  \bibinfo{author}{\bibfnamefont{O.}~\bibnamefont{Friedt}},
  \bibinfo{author}{\bibfnamefont{P.}~\bibnamefont{Alireza}},
  \bibinfo{author}{\bibfnamefont{W.~G.} \bibnamefont{Marshall}},
  \bibinfo{author}{\bibfnamefont{W.}~\bibnamefont{Schmidt}},
  \bibinfo{author}{\bibfnamefont{F.}~\bibnamefont{Nakamura}},
  \bibinfo{author}{\bibfnamefont{S.}~\bibnamefont{Nakatsuji}},
  \bibinfo{author}{\bibfnamefont{Y.}~\bibnamefont{Maeno}},
  \bibinfo{author}{\bibfnamefont{R.}~\bibnamefont{Lengsdorf}},
  \bibinfo{author}{\bibfnamefont{M.~M.} \bibnamefont{Abd-Elmeguid}},
  \bibnamefont{et~al.}, \bibinfo{journal}{Physical Review B}
  \textbf{\bibinfo{volume}{72}}, \bibinfo{pages}{094104}
  (\bibinfo{year}{2005}).

\bibitem[{\citenamefont{Friedt et~al.}(2001)\citenamefont{Friedt, Braden,
  Andre, Adelmann, Nakatsuji, and Maeno}}]{Friedt}
\bibinfo{author}{\bibfnamefont{O.}~\bibnamefont{Friedt}},
  \bibinfo{author}{\bibfnamefont{M.}~\bibnamefont{Braden}},
  \bibinfo{author}{\bibfnamefont{G.}~\bibnamefont{Andre}},
  \bibinfo{author}{\bibfnamefont{P.}~\bibnamefont{Adelmann}},
  \bibinfo{author}{\bibfnamefont{S.}~\bibnamefont{Nakatsuji}},
  \bibnamefont{and} \bibinfo{author}{\bibfnamefont{Y.}~\bibnamefont{Maeno}},
  \bibinfo{journal}{Physical Review B} \textbf{\bibinfo{volume}{63}},
  \bibinfo{pages}{174432} (\bibinfo{year}{2001}).

\bibitem[{\citenamefont{Pincini et~al.}(2018)\citenamefont{Pincini, Boseggia,
  Perry, Gutmann, Ricco, Veiga, Dashwood, Collins, Nisbet, Bombardi
  et~al.}}]{Pincini18}
\bibinfo{author}{\bibfnamefont{D.}~\bibnamefont{Pincini}},
  \bibinfo{author}{\bibfnamefont{S.}~\bibnamefont{Boseggia}},
  \bibinfo{author}{\bibfnamefont{R.}~\bibnamefont{Perry}},
  \bibinfo{author}{\bibfnamefont{M.~J.} \bibnamefont{Gutmann}},
  \bibinfo{author}{\bibfnamefont{S.}~\bibnamefont{Ricco}},
  \bibinfo{author}{\bibfnamefont{L.~S.~I.} \bibnamefont{Veiga}},
  \bibinfo{author}{\bibfnamefont{C.~D.} \bibnamefont{Dashwood}},
  \bibinfo{author}{\bibfnamefont{S.~P.} \bibnamefont{Collins}},
  \bibinfo{author}{\bibfnamefont{G.}~\bibnamefont{Nisbet}},
  \bibinfo{author}{\bibfnamefont{A.}~\bibnamefont{Bombardi}},
  \bibnamefont{et~al.}, \bibinfo{journal}{Physical Review B}
  \textbf{\bibinfo{volume}{98}}, \bibinfo{pages}{014429}
  (\bibinfo{year}{2018}).

\bibitem[{\citenamefont{Jung et~al.}(2003)\citenamefont{Jung, Fang, He, Kaneko,
  Okimoto, and Tokura}}]{Jung}
\bibinfo{author}{\bibfnamefont{J.~H.} \bibnamefont{Jung}},
  \bibinfo{author}{\bibfnamefont{Z.}~\bibnamefont{Fang}},
  \bibinfo{author}{\bibfnamefont{J.~P.} \bibnamefont{He}},
  \bibinfo{author}{\bibfnamefont{Y.}~\bibnamefont{Kaneko}},
  \bibinfo{author}{\bibfnamefont{Y.}~\bibnamefont{Okimoto}}, \bibnamefont{and}
  \bibinfo{author}{\bibfnamefont{Y.}~\bibnamefont{Tokura}},
  \bibinfo{journal}{Physical Review Letters} \textbf{\bibinfo{volume}{91}},
  \bibinfo{pages}{056403} (\bibinfo{year}{2003}), ISSN
  \bibinfo{issn}{0031-9007}.

\bibitem[{\citenamefont{Snow et~al.}(2002)\citenamefont{Snow, Cooper, Cao,
  Crow, Fukazawa, Nakatsuji, and Maeno}}]{Snow}
\bibinfo{author}{\bibfnamefont{C.~S.} \bibnamefont{Snow}},
  \bibinfo{author}{\bibfnamefont{S.~L.} \bibnamefont{Cooper}},
  \bibinfo{author}{\bibfnamefont{G.}~\bibnamefont{Cao}},
  \bibinfo{author}{\bibfnamefont{J.~E.} \bibnamefont{Crow}},
  \bibinfo{author}{\bibfnamefont{H.}~\bibnamefont{Fukazawa}},
  \bibinfo{author}{\bibfnamefont{S.}~\bibnamefont{Nakatsuji}},
  \bibnamefont{and} \bibinfo{author}{\bibfnamefont{Y.}~\bibnamefont{Maeno}},
  \bibinfo{journal}{Physical Review Letters} \textbf{\bibinfo{volume}{89}},
  \bibinfo{pages}{226401} (\bibinfo{year}{2002}), ISSN
  \bibinfo{issn}{0031-9007}.

\bibitem[{\citenamefont{Lee et~al.}(2002)\citenamefont{Lee, Lee, Noh, Oh, Yu,
  Nakatsuji, Fukazawa, and Maeno}}]{Lee02}
\bibinfo{author}{\bibfnamefont{J.~S.} \bibnamefont{Lee}},
  \bibinfo{author}{\bibfnamefont{Y.~S.} \bibnamefont{Lee}},
  \bibinfo{author}{\bibfnamefont{T.~W.} \bibnamefont{Noh}},
  \bibinfo{author}{\bibfnamefont{S.~J.} \bibnamefont{Oh}},
  \bibinfo{author}{\bibfnamefont{J.~J.} \bibnamefont{Yu}},
  \bibinfo{author}{\bibfnamefont{S.}~\bibnamefont{Nakatsuji}},
  \bibinfo{author}{\bibfnamefont{H.}~\bibnamefont{Fukazawa}}, \bibnamefont{and}
  \bibinfo{author}{\bibfnamefont{Y.}~\bibnamefont{Maeno}},
  \bibinfo{journal}{Physical Review Letters} \textbf{\bibinfo{volume}{89}},
  \bibinfo{pages}{257402} (\bibinfo{year}{2002}), ISSN
  \bibinfo{issn}{0031-9007}.

\bibitem[{\citenamefont{Qi et~al.}(2011)\citenamefont{Qi, Ge, Korneta, Parkin,
  De~Long, and Cao}}]{Qi_Cr}
\bibinfo{author}{\bibfnamefont{T.~F.} \bibnamefont{Qi}},
  \bibinfo{author}{\bibfnamefont{M.}~\bibnamefont{Ge}},
  \bibinfo{author}{\bibfnamefont{O.~B.} \bibnamefont{Korneta}},
  \bibinfo{author}{\bibfnamefont{S.}~\bibnamefont{Parkin}},
  \bibinfo{author}{\bibfnamefont{L.~E.} \bibnamefont{De~Long}},
  \bibnamefont{and} \bibinfo{author}{\bibfnamefont{G.}~\bibnamefont{Cao}},
  \bibinfo{journal}{Journal of Solid State Chemistry}
  \textbf{\bibinfo{volume}{184}}, \bibinfo{pages}{893} (\bibinfo{year}{2011}).

\bibitem[{\citenamefont{Qi et~al.}(2012)\citenamefont{Qi, Korneta, Parkin, Hu,
  and Cao}}]{Qi_Fe}
\bibinfo{author}{\bibfnamefont{T.~F.} \bibnamefont{Qi}},
  \bibinfo{author}{\bibfnamefont{O.~B.} \bibnamefont{Korneta}},
  \bibinfo{author}{\bibfnamefont{S.}~\bibnamefont{Parkin}},
  \bibinfo{author}{\bibfnamefont{J.~P.} \bibnamefont{Hu}}, \bibnamefont{and}
  \bibinfo{author}{\bibfnamefont{G.}~\bibnamefont{Cao}},
  \bibinfo{journal}{Physical Review B} \textbf{\bibinfo{volume}{85}},
  \bibinfo{pages}{165143} (\bibinfo{year}{2012}).

\bibitem[{\citenamefont{Liu and Khaliullin}(2019)}]{Liu19}
\bibinfo{author}{\bibfnamefont{H.~M.} \bibnamefont{Liu}} \bibnamefont{and}
  \bibinfo{author}{\bibfnamefont{G.}~\bibnamefont{Khaliullin}},
  \bibinfo{journal}{Physical Review Letters} \textbf{\bibinfo{volume}{122}},
  \bibinfo{pages}{057203} (\bibinfo{year}{2019}).

\bibitem[{\citenamefont{Lee et~al.}(2019)\citenamefont{Lee, Kim, Kwak, Seo,
  Sohn, Nakamura, Sow, Maeno, Kim, Noh et~al.}}]{Lee19}
\bibinfo{author}{\bibfnamefont{M.~C.} \bibnamefont{Lee}},
  \bibinfo{author}{\bibfnamefont{C.~H.} \bibnamefont{Kim}},
  \bibinfo{author}{\bibfnamefont{I.}~\bibnamefont{Kwak}},
  \bibinfo{author}{\bibfnamefont{C.~W.} \bibnamefont{Seo}},
  \bibinfo{author}{\bibfnamefont{C.}~\bibnamefont{Sohn}},
  \bibinfo{author}{\bibfnamefont{F.}~\bibnamefont{Nakamura}},
  \bibinfo{author}{\bibfnamefont{C.}~\bibnamefont{Sow}},
  \bibinfo{author}{\bibfnamefont{Y.}~\bibnamefont{Maeno}},
  \bibinfo{author}{\bibfnamefont{E.~A.} \bibnamefont{Kim}},
  \bibinfo{author}{\bibfnamefont{T.~W.} \bibnamefont{Noh}},
  \bibnamefont{et~al.}, \bibinfo{journal}{Physical Review B}
  \textbf{\bibinfo{volume}{99}}, \bibinfo{pages}{144306}
  (\bibinfo{year}{2019}).

\bibitem[{\citenamefont{Sheldrick}(2015)}]{SHELXT}
\bibinfo{author}{\bibfnamefont{G.~M.} \bibnamefont{Sheldrick}},
  \bibinfo{journal}{Acta Crystallographica a-Foundation and Advances}
  \textbf{\bibinfo{volume}{71}}, \bibinfo{pages}{3} (\bibinfo{year}{2015}).

\bibitem[{\citenamefont{Farrugia}(2012)}]{Farrugia}
\bibinfo{author}{\bibfnamefont{L.~J.} \bibnamefont{Farrugia}},
  \bibinfo{journal}{Journal of Applied Crystallography}
  \textbf{\bibinfo{volume}{45}}, \bibinfo{pages}{849} (\bibinfo{year}{2012}).

\bibitem[{\citenamefont{Burla et~al.}(2012)\citenamefont{Burla, Caliandro,
  Camalli, Carrozzini, Cascarano, Giacovazzo, Mallamo, Mazzone, Polidori, and
  Spagna}}]{Burla}
\bibinfo{author}{\bibfnamefont{M.~C.} \bibnamefont{Burla}},
  \bibinfo{author}{\bibfnamefont{R.}~\bibnamefont{Caliandro}},
  \bibinfo{author}{\bibfnamefont{M.}~\bibnamefont{Camalli}},
  \bibinfo{author}{\bibfnamefont{B.}~\bibnamefont{Carrozzini}},
  \bibinfo{author}{\bibfnamefont{G.~L.} \bibnamefont{Cascarano}},
  \bibinfo{author}{\bibfnamefont{C.}~\bibnamefont{Giacovazzo}},
  \bibinfo{author}{\bibfnamefont{M.}~\bibnamefont{Mallamo}},
  \bibinfo{author}{\bibfnamefont{A.}~\bibnamefont{Mazzone}},
  \bibinfo{author}{\bibfnamefont{G.}~\bibnamefont{Polidori}}, \bibnamefont{and}
  \bibinfo{author}{\bibfnamefont{R.}~\bibnamefont{Spagna}},
  \bibinfo{journal}{Journal of Applied Crystallography}
  \textbf{\bibinfo{volume}{45}}, \bibinfo{pages}{357} (\bibinfo{year}{2012}).

\bibitem[{\citenamefont{Rodriguezcarvajal}(1993)}]{FullProf}
\bibinfo{author}{\bibfnamefont{J.}~\bibnamefont{Rodriguezcarvajal}},
  \bibinfo{journal}{Physica B} \textbf{\bibinfo{volume}{192}},
  \bibinfo{pages}{55} (\bibinfo{year}{1993}).

\bibitem[{\citenamefont{Yuan et~al.}(2016)\citenamefont{Yuan, Qi, Terzic,
  Zheng, Zhao, Chi, Ye, Wei, Parkin, Liu et~al.}}]{Yuan}
\bibinfo{author}{\bibfnamefont{S.~J.} \bibnamefont{Yuan}},
  \bibinfo{author}{\bibfnamefont{T.~F.} \bibnamefont{Qi}},
  \bibinfo{author}{\bibfnamefont{J.}~\bibnamefont{Terzic}},
  \bibinfo{author}{\bibfnamefont{H.}~\bibnamefont{Zheng}},
  \bibinfo{author}{\bibfnamefont{Z.}~\bibnamefont{Zhao}},
  \bibinfo{author}{\bibfnamefont{S.}~\bibnamefont{Chi}},
  \bibinfo{author}{\bibfnamefont{F.}~\bibnamefont{Ye}},
  \bibinfo{author}{\bibfnamefont{H.}~\bibnamefont{Wei}},
  \bibinfo{author}{\bibfnamefont{S.}~\bibnamefont{Parkin}},
  \bibinfo{author}{\bibfnamefont{X.}~\bibnamefont{Liu}}, \bibnamefont{et~al.},
  \emph{\bibinfo{title}{Magnetization reversal and negative volume thermal
  expansion in fe doped ca2ruo4}} (\bibinfo{year}{2016}),
  \eprint{arXiv:1605.06352}.

\bibitem[{\citenamefont{Blume}(1961)}]{Blume}
\bibinfo{author}{\bibfnamefont{M.}~\bibnamefont{Blume}},
  \bibinfo{journal}{Physical Review} \textbf{\bibinfo{volume}{124}},
  \bibinfo{pages}{96} (\bibinfo{year}{1961}).

\bibitem[{\citenamefont{Belov et~al.}(1957)\citenamefont{Belov, Neronova, and
  Smirnov}}]{BNS}
\bibinfo{author}{\bibfnamefont{N.~V.} \bibnamefont{Belov}},
  \bibinfo{author}{\bibfnamefont{N.~N.} \bibnamefont{Neronova}},
  \bibnamefont{and} \bibinfo{author}{\bibfnamefont{T.~S.}
  \bibnamefont{Smirnov}}, \bibinfo{journal}{Kristallografiya}
  \textbf{\bibinfo{volume}{2}}, \bibinfo{pages}{315} (\bibinfo{year}{1957}).

\bibitem[{\citenamefont{Opechowski and Guccione}(1965)}]{OG}
\bibinfo{author}{\bibnamefont{Opechowski}} \bibnamefont{and}
  \bibinfo{author}{\bibnamefont{Guccione}}, \emph{\bibinfo{title}{Magnetism}}
  (\bibinfo{publisher}{New York: Academic Press}, \bibinfo{year}{1965}).

\bibitem[{\citenamefont{Cuoco et~al.}(2006{\natexlab{a}})\citenamefont{Cuoco,
  Forte, and Noce}}]{Cuoco2}
\bibinfo{author}{\bibfnamefont{M.}~\bibnamefont{Cuoco}},
  \bibinfo{author}{\bibfnamefont{F.}~\bibnamefont{Forte}}, \bibnamefont{and}
  \bibinfo{author}{\bibfnamefont{C.}~\bibnamefont{Noce}},
  \bibinfo{journal}{Physical Review B} \textbf{\bibinfo{volume}{73}},
  \bibinfo{pages}{094428} (\bibinfo{year}{2006}{\natexlab{a}}).

\bibitem[{\citenamefont{Rams et~al.}(2009)\citenamefont{Rams, Kruzel, Zarzycki,
  Krolas, and Tomala}}]{Rams}
\bibinfo{author}{\bibfnamefont{M.}~\bibnamefont{Rams}},
  \bibinfo{author}{\bibfnamefont{M.}~\bibnamefont{Kruzel}},
  \bibinfo{author}{\bibfnamefont{A.}~\bibnamefont{Zarzycki}},
  \bibinfo{author}{\bibfnamefont{K.}~\bibnamefont{Krolas}}, \bibnamefont{and}
  \bibinfo{author}{\bibfnamefont{K.}~\bibnamefont{Tomala}},
  \bibinfo{journal}{Physical Review B} \textbf{\bibinfo{volume}{80}},
  \bibinfo{pages}{045119} (\bibinfo{year}{2009}).

\bibitem[{\citenamefont{Endoh et~al.}(1988)\citenamefont{Endoh, Yamada,
  Birgeneau, Gabbe, Jenssen, Kastner, Peters, Picone, Thurston, Tranquada
  et~al.}}]{Endoh}
\bibinfo{author}{\bibfnamefont{Y.}~\bibnamefont{Endoh}},
  \bibinfo{author}{\bibfnamefont{K.}~\bibnamefont{Yamada}},
  \bibinfo{author}{\bibfnamefont{R.~J.} \bibnamefont{Birgeneau}},
  \bibinfo{author}{\bibfnamefont{D.~R.} \bibnamefont{Gabbe}},
  \bibinfo{author}{\bibfnamefont{H.~P.} \bibnamefont{Jenssen}},
  \bibinfo{author}{\bibfnamefont{M.~A.} \bibnamefont{Kastner}},
  \bibinfo{author}{\bibfnamefont{C.~J.} \bibnamefont{Peters}},
  \bibinfo{author}{\bibfnamefont{P.~J.} \bibnamefont{Picone}},
  \bibinfo{author}{\bibfnamefont{T.~R.} \bibnamefont{Thurston}},
  \bibinfo{author}{\bibfnamefont{J.~M.} \bibnamefont{Tranquada}},
  \bibnamefont{et~al.}, \bibinfo{journal}{Physical Review B}
  \textbf{\bibinfo{volume}{37}}, \bibinfo{pages}{7443} (\bibinfo{year}{1988}).

\bibitem[{\citenamefont{Wilson et~al.}(2010)\citenamefont{Wilson, Yamani,
  Rotundu, Freelon, Valdivia, Bourret-Courchesne, Lynn, Chi, Hong, and
  Birgeneau}}]{Wilson}
\bibinfo{author}{\bibfnamefont{S.~D.} \bibnamefont{Wilson}},
  \bibinfo{author}{\bibfnamefont{Z.}~\bibnamefont{Yamani}},
  \bibinfo{author}{\bibfnamefont{C.~R.} \bibnamefont{Rotundu}},
  \bibinfo{author}{\bibfnamefont{B.}~\bibnamefont{Freelon}},
  \bibinfo{author}{\bibfnamefont{P.~N.} \bibnamefont{Valdivia}},
  \bibinfo{author}{\bibfnamefont{E.}~\bibnamefont{Bourret-Courchesne}},
  \bibinfo{author}{\bibfnamefont{J.~W.} \bibnamefont{Lynn}},
  \bibinfo{author}{\bibfnamefont{S.~X.} \bibnamefont{Chi}},
  \bibinfo{author}{\bibfnamefont{T.}~\bibnamefont{Hong}}, \bibnamefont{and}
  \bibinfo{author}{\bibfnamefont{R.~J.} \bibnamefont{Birgeneau}},
  \bibinfo{journal}{Physical Review B} \textbf{\bibinfo{volume}{82}},
  \bibinfo{pages}{144502} (\bibinfo{year}{2010}).

\bibitem[{\citenamefont{Birgeneau et~al.}(1971)\citenamefont{Birgeneau, Skalyo,
  and Shirane}}]{Birgeneau}
\bibinfo{author}{\bibfnamefont{R.~J.} \bibnamefont{Birgeneau}},
  \bibinfo{author}{\bibfnamefont{J.}~\bibnamefont{Skalyo}}, \bibnamefont{and}
  \bibinfo{author}{\bibfnamefont{G.}~\bibnamefont{Shirane}},
  \bibinfo{journal}{Physical Review B-Solid State}
  \textbf{\bibinfo{volume}{3}}, \bibinfo{pages}{1736} (\bibinfo{year}{1971}).

\bibitem[{\citenamefont{Wollan and Koehler}(1955)}]{Wollan}
\bibinfo{author}{\bibfnamefont{E.~O.} \bibnamefont{Wollan}} \bibnamefont{and}
  \bibinfo{author}{\bibfnamefont{W.~C.} \bibnamefont{Koehler}},
  \bibinfo{journal}{Physical Review} \textbf{\bibinfo{volume}{100}},
  \bibinfo{pages}{545} (\bibinfo{year}{1955}).

\bibitem[{\citenamefont{Fang and Terakura}(2001)}]{Fang}
\bibinfo{author}{\bibfnamefont{Z.}~\bibnamefont{Fang}} \bibnamefont{and}
  \bibinfo{author}{\bibfnamefont{K.}~\bibnamefont{Terakura}},
  \bibinfo{journal}{Physical Review B} \textbf{\bibinfo{volume}{64}},
  \bibinfo{pages}{020509(R)} (\bibinfo{year}{2001}).

\bibitem[{\citenamefont{Taniguchi et~al.}(2013)\citenamefont{Taniguchi,
  Nishimura, Ishikawa, Yonezawa, Goh, Nakamura, and Maeno}}]{Taniguchi}
\bibinfo{author}{\bibfnamefont{H.}~\bibnamefont{Taniguchi}},
  \bibinfo{author}{\bibfnamefont{K.}~\bibnamefont{Nishimura}},
  \bibinfo{author}{\bibfnamefont{R.}~\bibnamefont{Ishikawa}},
  \bibinfo{author}{\bibfnamefont{S.}~\bibnamefont{Yonezawa}},
  \bibinfo{author}{\bibfnamefont{S.~K.} \bibnamefont{Goh}},
  \bibinfo{author}{\bibfnamefont{F.}~\bibnamefont{Nakamura}}, \bibnamefont{and}
  \bibinfo{author}{\bibfnamefont{Y.}~\bibnamefont{Maeno}},
  \bibinfo{journal}{Physical Review B} \textbf{\bibinfo{volume}{88}},
  \bibinfo{pages}{205111} (\bibinfo{year}{2013}).

\bibitem[{\citenamefont{Carlo et~al.}(2012)\citenamefont{Carlo, Goko,
  Gat-Malureanu, Russo, Savici, Aczel, MacDougall, Rodriguez, Williams, Luke
  et~al.}}]{Carlo}
\bibinfo{author}{\bibfnamefont{J.~P.} \bibnamefont{Carlo}},
  \bibinfo{author}{\bibfnamefont{T.}~\bibnamefont{Goko}},
  \bibinfo{author}{\bibfnamefont{I.~M.} \bibnamefont{Gat-Malureanu}},
  \bibinfo{author}{\bibfnamefont{P.~L.} \bibnamefont{Russo}},
  \bibinfo{author}{\bibfnamefont{A.~T.} \bibnamefont{Savici}},
  \bibinfo{author}{\bibfnamefont{A.~A.} \bibnamefont{Aczel}},
  \bibinfo{author}{\bibfnamefont{G.~J.} \bibnamefont{MacDougall}},
  \bibinfo{author}{\bibfnamefont{J.~A.} \bibnamefont{Rodriguez}},
  \bibinfo{author}{\bibfnamefont{T.~J.} \bibnamefont{Williams}},
  \bibinfo{author}{\bibfnamefont{G.~M.} \bibnamefont{Luke}},
  \bibnamefont{et~al.}, \bibinfo{journal}{Nature Materials}
  \textbf{\bibinfo{volume}{11}}, \bibinfo{pages}{323} (\bibinfo{year}{2012}),
  ISSN \bibinfo{issn}{1476-1122}.

\bibitem[{\citenamefont{Pincini et~al.}(2019)\citenamefont{Pincini, Veiga,
  Dashwood, Forte, Cuoco, Perry, Bencok, Boothroyd, and McMorrow}}]{Pincini19}
\bibinfo{author}{\bibfnamefont{D.}~\bibnamefont{Pincini}},
  \bibinfo{author}{\bibfnamefont{L.~S.~I.} \bibnamefont{Veiga}},
  \bibinfo{author}{\bibfnamefont{C.~D.} \bibnamefont{Dashwood}},
  \bibinfo{author}{\bibfnamefont{F.}~\bibnamefont{Forte}},
  \bibinfo{author}{\bibfnamefont{M.}~\bibnamefont{Cuoco}},
  \bibinfo{author}{\bibfnamefont{R.~S.} \bibnamefont{Perry}},
  \bibinfo{author}{\bibfnamefont{P.}~\bibnamefont{Bencok}},
  \bibinfo{author}{\bibfnamefont{A.~T.} \bibnamefont{Boothroyd}},
  \bibnamefont{and} \bibinfo{author}{\bibfnamefont{D.~F.}
  \bibnamefont{McMorrow}}, \bibinfo{journal}{Physical Review B}
  \textbf{\bibinfo{volume}{99}}, \bibinfo{pages}{075125}
  (\bibinfo{year}{2019}).

\bibitem[{\citenamefont{Cuoco et~al.}(2006{\natexlab{b}})\citenamefont{Cuoco,
  Forte, and Noce}}]{Cuoco1}
\bibinfo{author}{\bibfnamefont{M.}~\bibnamefont{Cuoco}},
  \bibinfo{author}{\bibfnamefont{F.}~\bibnamefont{Forte}}, \bibnamefont{and}
  \bibinfo{author}{\bibfnamefont{C.}~\bibnamefont{Noce}},
  \bibinfo{journal}{Physical Review B} \textbf{\bibinfo{volume}{74}},
  \bibinfo{pages}{195124} (\bibinfo{year}{2006}{\natexlab{b}}).

\bibitem[{\citenamefont{Nakamura et~al.}(2002)\citenamefont{Nakamura, Goko,
  Ito, Fujita, Nakatsuji, Fukazawa, Maeno, Alireza, Forsythe, and
  Julian}}]{Nakamura}
\bibinfo{author}{\bibfnamefont{F.}~\bibnamefont{Nakamura}},
  \bibinfo{author}{\bibfnamefont{T.}~\bibnamefont{Goko}},
  \bibinfo{author}{\bibfnamefont{M.}~\bibnamefont{Ito}},
  \bibinfo{author}{\bibfnamefont{T.}~\bibnamefont{Fujita}},
  \bibinfo{author}{\bibfnamefont{S.}~\bibnamefont{Nakatsuji}},
  \bibinfo{author}{\bibfnamefont{H.}~\bibnamefont{Fukazawa}},
  \bibinfo{author}{\bibfnamefont{Y.}~\bibnamefont{Maeno}},
  \bibinfo{author}{\bibfnamefont{P.}~\bibnamefont{Alireza}},
  \bibinfo{author}{\bibfnamefont{D.}~\bibnamefont{Forsythe}}, \bibnamefont{and}
  \bibinfo{author}{\bibfnamefont{S.~R.} \bibnamefont{Julian}},
  \bibinfo{journal}{Physical Review B} \textbf{\bibinfo{volume}{65}},
  \bibinfo{pages}{220402(R)} (\bibinfo{year}{2002}).

\bibitem[{\citenamefont{Fukazawa and Maeno}(2000)}]{Fukazawa}
\bibinfo{author}{\bibfnamefont{H.}~\bibnamefont{Fukazawa}} \bibnamefont{and}
  \bibinfo{author}{\bibfnamefont{Y.}~\bibnamefont{Maeno}},
  \bibinfo{journal}{Journal of the Physical Society of Japan}
  \textbf{\bibinfo{volume}{70}}, \bibinfo{pages}{460} (\bibinfo{year}{2000}).

\bibitem[{\citenamefont{Porter et~al.}(2018)\citenamefont{Porter, Granata,
  Forte, Di~Matteo, Cuoco, Fittipaldi, Vecchione, and Bombardi}}]{Porter}
\bibinfo{author}{\bibfnamefont{D.~G.} \bibnamefont{Porter}},
  \bibinfo{author}{\bibfnamefont{V.}~\bibnamefont{Granata}},
  \bibinfo{author}{\bibfnamefont{F.}~\bibnamefont{Forte}},
  \bibinfo{author}{\bibfnamefont{S.}~\bibnamefont{Di~Matteo}},
  \bibinfo{author}{\bibfnamefont{M.}~\bibnamefont{Cuoco}},
  \bibinfo{author}{\bibfnamefont{R.}~\bibnamefont{Fittipaldi}},
  \bibinfo{author}{\bibfnamefont{A.}~\bibnamefont{Vecchione}},
  \bibnamefont{and} \bibinfo{author}{\bibfnamefont{A.}~\bibnamefont{Bombardi}},
  \bibinfo{journal}{Physical Review B} \textbf{\bibinfo{volume}{98}},
  \bibinfo{pages}{125142} (\bibinfo{year}{2018}).

\end{thebibliography}
\end{document}